\documentclass[]{aa}  
\usepackage{subfigure}
\usepackage{graphicx}
\usepackage{xcolor}
\usepackage{txfonts}
\usepackage{natbib,twoopt}
\usepackage[breaklinks=true]{hyperref} 
\bibpunct{(}{)}{;}{a}{}{,}             
\makeatletter
  \newcommandtwoopt{\citeads}[3][][]{\href{http://adsabs.harvard.edu/abs/#3}%
    {\def\hyper@linkstart##1##2{}%
     \let\hyper@linkend\@empty\citealp[#1][#2]{#3}}}
  \newcommandtwoopt{\citepads}[3][][]{\href{http://adsabs.harvard.edu/abs/#3}%
    {\def\hyper@linkstart##1##2{}%
     \let\hyper@linkend\@empty\citep[#1][#2]{#3}}}
  \newcommandtwoopt{\citetads}[3][][]{\href{http://adsabs.harvard.edu/abs/#3}%
    {\def\hyper@linkstart##1##2{}%
     \let\hyper@linkend\@empty\citet[#1][#2]{#3}}}
  \newcommandtwoopt{\citeyearads}[3][][]%
    {\href{http://adsabs.harvard.edu/abs/#3}
    {\def\hyper@linkstart##1##2{}%
     \let\hyper@linkend\@empty\citeyear[#1][#2]{#3}}}
\makeatother

\begin{document}

   \title{The perils of stacking optically selected groups in eROSITA data. \\
   The Magneticum perspective}


   \author{P. Popesso\inst{1,}\inst{2}\thanks{paola.popesso@eso.org}
   \and
          I. Marini\inst{1}
          \and
          K. Dolag\inst{3,}\inst{4,}\inst{2}
          \and
          G. Lamer\inst{5}
          \and
          B. Csizi\inst{6}
          \and
          S. Vladutescu-Zopp\inst{3}
          \and
          V. Biffi\inst{7}
        \and
        A. Robothan\inst{8}
        \and 
        M. Bravo\inst{9}
          \and
          E. Tempel\inst{10}
          \and
          X. Yang\inst{11}
          \and
          Q. Li\inst{11}
          \and
          A. Biviano\inst{7,}\inst{12}
          \and
          L. Lovisari\inst{13,}\inst{14,}
          \and
          S. Ettori\inst{15}
          \and
          M. Angelinelli\inst{15}
          \and
          S. Driver\inst{8}
          \and
          V. Toptun\inst{1}
          \and
          A. Dev\inst{8}
          \and
          D. Mazengo\inst{1}
          \and
          A. Merloni\inst{16}
          \and
          T. Mroczkowski\inst{1}
          \and 
          J. Comparat\inst{16}
          \and
          Y. Zhang\inst{16}
          \and
          G. Ponti\inst{16}
          \and
          E. Bulbul\inst{16}
          }

   \institute{European Southern Observatory, Karl Schwarzschildstrasse 2, 85748, Garching bei M\"unchen, Germany \email{paola.popesso@eso.org}
         \and
            Excellence Cluster ORIGINS, Boltzmannstr. 2, D-85748 Garching bei M\"unchen, Germany
        \and
             Universitäts-Sternwarte, Fakultät für Physik, Ludwig-Maximilians-Universität München, Scheinerstr.1, 81679 München, Germany
        \and 
            Max-Planck-Institut für Astrophysik, Karl-Schwarzschildstr. 1, 85741 Garching bei M\"unchen, Germany
        \and
            Leibniz-Institut für Astrophysik Potsdam (AIP), An der Sternwarte 16, 14482 Potsdam, Germany   
        \and
        Universität Innsbruck, Institut für Astro- und Teilchenphysik, Technikerstr. 25/8, 6020 Innsbruck, Austria
        \and
             INAF – Osservatorio Astronomico di Trieste, Via Tiepolo 11, 34143 Trieste, Italy
            \and
            International Centre for Radio Astronomy Research, University of Western Australia, M468, 35 Stirling Highway, Perth, WA 6009, Australia
            \and 
            McMaster University, 1280 Main Street West, Hamilton, Ontario, L8S 4L8, Canada
            \and
            Tartu University, Ülikooli 18, 50090 Tartu, Estonia
            \and
            Shanghai Astronomical Observatory (SHAO) at the Chinese Academy of Sciences, 80 Nandan Road, Xuhui District, Shanghai 200030, China
        \and 
            IFPU – Institute for Fundamental Physics of the Universe, Via Beirut 2, I-34014 Trieste, Italy
        \and
            INAF– Osservatorio Astronomico di Brera, Via E. Bianchi 46, 23807 Merate (LC), Italy
                        \and
            Center for Astrophysics $|$ Harvard $\&$ Smithsonian, 60 Garden Street, Cambridge, MA 02138, USA
        \and
            INAF– Osservatorio Astronomico di Bologna, Via Gobetti 93/3, 40129 Bologna, Italy
                \and
            Max-Planck-Institut für Extraterrestrische Physik (MPE), Giessenbachstr. 1, D-85748 Garching bei München, Germany
             }

   \date{Received September 15, 1996; accepted March 16, 1997}

 
  \abstract
   {Hydrodynamical simulation predictions are often compared with observational data without fully accounting for systematics and biases specific to observational techniques. In this study, we use the magnetohydrodynamical simulation {\it{Magneticum}} to create a mock dataset that replicates the observational data available for analyzing hot gas properties in extensive galaxy group samples. }
   {Specifically, we simulate eROSITA eRASS:4 data along with a GAMA-like galaxy spectroscopic survey from the same lightcone and generate mock, optically selected galaxy catalogs using widely employed group-finding algorithms. We then apply an observational stacking technique to the mock eRASS:4 observations, using the mock group catalogs as priors to determine the average properties of the underlying group population. This approach serves two primary purposes: (i) to produce predictions that incorporate observational systematics, and (ii) to assess these systematics and evaluate the reliability of the stacking technique in deriving the average X-ray properties of galaxy groups from eROSITA data.}
   {We provide the predicted X-ray emission of the Magneticum simulation divided into all contributions of AGN, X-ray binaries (XRB), and Intra-Group Medium (IGM) per bin of halo mass. The predicted AGN and XRB contamination dominates the X-ray surface brightness profile emission in all halos with masses below $10^{13}$ $M_{\odot}$, which contains the majority of low X-ray luminosity AGN. We test the reliability of the stacking technique in reproducing the input X-ray surface brightness and electron density profile for all tested optical group selection algorithms. We consider completeness and contamination of the prior samples, mis-centering of the optical group center, uncertainties in determining the X-ray emissivity due to the assumptions of mean gas temperature and metallicity, and systematics in the available halo mass proxy.}
   {The primary source of systematics in our analysis arises from the precision of the halo mass proxy, which might impact the estimation of X-ray surface brightness profiles when utilized as a prior, and derived scaling relations. Our analysis displays the $L_X-$mass relationships produced by stacking various optically selected group priors and reveals that, in each instance, the slope of these relations appears somewhat flatter than the input relation, though still in agreement with observational data. We retrieve the $f_{gas}$-mass relation within $R_{500}$ effectively and find agreement between the predictions and observational data.}
   {Such systematic errors must be considered when comparing the results of any stacking technique with other works in the literature based on different prior catalogs, detections, or predictions.}


   \maketitle
%

\section{Introduction}

Galaxy groups represent the new frontier of modern cosmology, serving as crucial testbeds for predictions of contemporary models of large-scale structure and galaxy formation and evolution \citep{McCarthy2017}. The hot gas content, and consequently the baryonic mass and X-ray appearance of these groups, are significantly influenced by feedback from the supermassive black hole (BH) at the center of the central galaxy. At the cluster mass scale, such effects are typically confined to the central core \citep{LeBrun2014}. However, at the group mass scale, the energy involved is comparable to the system's binding energy, potentially impacting the entire group volume on a megaparsec scale \citep{Oppenheimer2020,2021Univ....7..142E}.

Modeling mechanisms such as AGN or stellar feedback, which expel baryons from collapsed structures and trigger baryonic exchange, is currently both the major strength and greatest challenge of our galaxy formation and evolution paradigm. Sophisticated numerical simulations have been developed to model the interplay between feedback and gas cooling in massive systems. However, predictions vary significantly regarding the amount and distribution of baryonic mass in groups, ranging from exceedingly hot gas-rich halos to systems entirely devoid of gas \citep{Oppenheimer2020}. These differences stem primarily from variations in feedback implementation. These range from a single feedback mode, where a percentage of BH energy is deposited into neighboring cells via a thermal bump (Eagle, \citep{Schaye2015}; BahamasXL, \citep{McCarthy2017}; and FLAMINGO, \citep{Schaye2023}), to more sophisticated two-mode implementations. These include thermal dumps or BH-driven outflows at high accretion rates (quasar mode) and kinetic kicks or energy dumping through bubbles at low accretion rates (radio mode), as seen in IllustrisTNG \citep{pillepich19} and Magneticum \citep{Dolag16}. Firm observational constraints on the X-ray appearance and hot gas content of galaxy groups are essential for navigating these potential solutions and predictions.

Until now, obtaining such observational constraints has been nearly impossible due to the lack of sensitivity in the soft X-ray energy bands or the survey capabilities of previous telescopes. eROSITA, with its high sensitivity below 2 keV and its all-sky survey \citep[eRASS:1, where the 1 indicates the first sky passage][]{Bulbul24}, will provide unprecedented statistics in the X-ray luminosity and halo mass regimes of galaxy groups. However, using the much deeper eFEDS data, \citet{Popesso24} show that the eROSITA X-ray selection (at the nominal eRASS:8 depth, at the completion of the survey) captures only a marginal fraction of the galaxy group population below halo masses of $10^{14}$ $M_{\odot}$. Additionally, using synthetic eROSITA data based on Magneticum hydrodynamical simulation light-cones, \citet{Marini2024a} demonstrate that the eROSITA selection function is biased against galaxy groups with lower surface brightness profiles and higher core entropy. This bias is expected if non-gravitational processes, such as AGN feedback, are strong enough to expel gas beyond the group's virial radius, thereby lowering the hot gas density and the group’s X-ray appearance and luminosity ($L_X$), which is proportional to the square of the gas density. This aligns with the eROSITA Consortium analysis of the eROSITA selection function of extended sources \citep{Biffi18, Clerc2018,Seppi2022}.

To circumvent such selection effects and study the average properties of the bulk of the group population, many works in the literature use the stacking technique. Stacking X-ray data of groups selected through different and less biased techniques can provide insights into the average properties of the population. The results of stacking are meaningful under the assumption that the scatter around the average value of the chosen property follows a single normal or lognormal distribution. Instead, for a dichotomic distribution, the stacking might lead to non-representative properties. This approach has been applied in the past, particularly by stacking optically selected galaxy group samples in the ROSAT data \citep[see e.g.][]{Anderson2015, rozo2009}. Groups are typically binned by optical properties that correlate with halo mass, such as the stellar mass of the central galaxy \citep{Anderson2015} or group and cluster richness \citep{rozo2009}. However, due to the large PSF, stacking on ROSAT data only provided a detection in the X-rays without much spatial information. The higher spatial resolution of eROSITA now allows for the estimation of the average X-ray surface brightness profile, which can infer useful information such as gas mass and spatial distribution. Recently, this technique has been used on eROSITA data to stack SDSS galaxy group catalogs \citep{Zhang24a} or GAMA-based catalogs \citep{Popesso24}. Nevertheless, the exact shape and flux level of the retrieved profiles remain uncertain due to the systematics of the stacking analysis, such as the completeness and contamination levels of the prior samples, contamination by other X-ray emitting sources such as AGN and X-ray binaries (XRB), and the accuracy of the prior properties for binning the groups.

To test these systematics and evaluate the reliability of the stacking technique in inferring the average properties of galaxy groups from eROSITA data, we generate synthetic observations that closely mimic the eROSITA eRASS:4 data \citep{Marini2024a} and optical galaxy catalogs \citep{Marini24b}. These synthetic datasets are constructed using dedicated lightcones from the magneto-hydrodynamical simulation {\it{Magneticum}} \citep{Dolag16}. By replicating the GAMA selection and employing group finders from \citet{Robotham2011}, \citet{Yang2005}, and \citet{Tempel2017}, we create synthetic samples of optically selected galaxy groups. These synthetic samples are then stacked in bins of the available halo mass proxy derived from the synthetic eROSITA observations, allowing us to assess the reliability of the stacking results.

As a byproduct, this experiment will provide predictions from the {\it{Magneticum}} simulation, both directly from the simulated data and after applying the instrumental and observational effects characteristic of eROSITA observations and optical group selection. In addition to a direct test of the observational technique by comparing input and output, this approach will yield a theoretical benchmark, facilitating a meaningful comparison with real observational data that incorporates the inherent systematics and uncertainties of the observations.

The paper is structured as follows. In Section 2, we describe the synthetic dataset composed of the galaxy catalogs and synthetic eROSITA observations. In Section 3, we provide the predicted X-ray emission of the Magneticum simulation divided into all contributions (AGN, XRB, and IGM) per bin of halo mass. In Section 4, we test the reliability of the stacking technique in reproducing the input X-ray surface brightness profile and, consequently, the electron density profiles. In Section 5, we compare Magneticum predictions with observational results based on the stacking technique applied to eROSITA data. In Section 6, we present the implications of the systematics of stacking in the analysis of the $L_X$-mass and $f_{gas}$-mass relations. Section 7 provides our discussion and conclusions.

Throughout the paper, we assume a flat $\Lambda$CDM cosmology with $H_0 = 67.74$ km~s$^{-1}$~Mpc$^{-1}$ and $\Omega{_m}(z = 0) = 0.3089$ \citep{Planck2016}.

\begin{figure}
\includegraphics[width=0.5\textwidth]{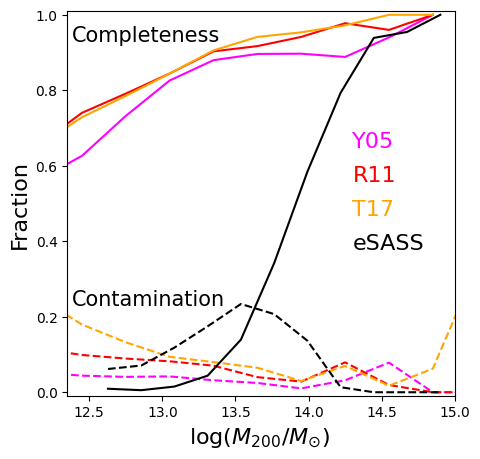}
\caption{Completeness (solid lines) and contamination (dashed lines) of the optical group catalogs based on the Magneticum mock galaxy catalog: R11 (red lines), T17 (orange lines), and Y05 (magenta lines) as obtained in \citet{Marini2024a}. The solid and dashed black lines indicate completeness and contamination, respectively, of the X-ray selected sample extracted from the eROSITA mock observation with the eSASS detection algorithm.}
\label{fig0}
\end{figure}

\section{The synthetic dataset}
To verify each step of the optical group selection and stacking procedure in the eROSITA data, we construct mock galaxy group catalogs based on different algorithms and a mock eROSITA observation from the same light-cone of the Magneticum simulation, creating analogs of standard observational datasets. A detailed description of the Magneticum lightcones and of the creation of the mock datasets is available in \citet{Marini2024a,Marini24b}. Here we provide a brief summary of the relevant aspects.

The Magneticum Pathfinder simulation\footnote{\url{http://www.magneticum.org/index.html}} is a comprehensive set of state-of-the-art cosmological hydrodynamical simulations conducted with the P-GADGET3 code \citep[][]{springel_cosmological_2003}. Significant improvements include a higher-order kernel function, time-dependent artificial viscosity, and artificial conduction schemes \citep{dolag_turbulent_2005,beck_improved_2016}. These simulations incorporate various subgrid models to account for unresolved baryonic physics, such as radiative cooling \citep{wiersma_effect_2009}, a uniform time-dependent UV background \citep{haardt_modelling_2001}, star formation and stellar feedback \citep[i.e., galactic winds;][]{springel_cosmological_2003}, and explicit chemical enrichment from stellar evolution \citep{tornatore_chemical_2007}. Additionally, they include models for supermassive black hole (SMBH) growth, accretion, and AGN feedback following established methodologies \citep{springel_cosmological_2003,di_matteo_energy_2005, fabjan_simulating_2010, hirschmann_cosmological_2014}.

The L30 lightcone derived from the Magneticum simulation includes the optical emission from galaxies and AGN, together with the X-ray emission from hot gas, AGN, and X-ray binaries for all sources in an area of 30$\times$30 deg$^{2}$ at $z <0.2$.  Detailed information on the generation of the mock eROSITA observations is provided in \citet{Marini2024a}, while the GAMA-like mock catalog and galaxy group sample are described in \citet{Marini24b}. Here, we offer a brief description of the synthetic optical and X-ray datasets to emphasize their similarities with the observed data.

\subsection{The GAMA-like mock galaxy catalog and galaxy group samples}
\label{optical_mock}
The galaxy mock catalog is derived from the light cones of the Magneticum simulation. The galaxy and halo catalogs within Magneticum are identified using the SubFind halo finder \citep{springel_populating_2001,dolag_substructures_2009}, which compiles a comprehensive list of observables (e.g., stellar mass, halo mass, star formation) by integrating the properties of the constituent particles. The mock galaxy catalog generated from the light-cone is limited to the local Universe up to $z < 0.2$ and covers an area of 30$\times$30 deg$^2$. It includes synthetic magnitudes in the SDSS filters (u, g, r, i, z), observed redshifts, stellar mass, and projected positions on the sky (i.e., RA, Dec) for each galaxy. To simulate a GAMA-like survey, we apply an r-band magnitude cut at 19.8 mag. To mimic observational uncertainties, observed redshifts and stellar masses are assigned errors drawn from Gaussian distributions with $\sigma = 45$ km s$^{-1}$ and 0.2 dex, respectively. Additionally, 5\% of the galaxies are set to undergo catastrophic failure in the spectroscopic survey (i.e., $\Delta v > 500$ km s$^{-1}$), and a spectroscopic completeness of 95\% is simulated.

To create optically selected galaxy group catalogs analogous to those available in the literature, we apply the galaxy group finder algorithm of \cite[][R11, hereafter]{Robotham2011}, applied to the GAMA survey, the \citet[][Y05, hereafer]{Yang2005} and \citet[][T17 hereafter]{Tempel2017} algorithms applied to the spectroscopic sample of the SDSS. These approaches use a Friends-of-Friends (FoF) algorithm for galaxy-galaxy linking, which has been extensively tested on synthetic mock catalogs in the corresponding reference paper. The individual algorithm performance has been thoroughly tested in \citet{Marini24b} by matching the input Magneticum halo catalog with the group catalog in terms of coordinates, redshift, and mass. Here, we report the main results regarding the algorithm's performance in terms of completeness, contamination, and the best halo mass proxy, as these are particularly relevant for the stacking analysis in the corresponding mock eROSITA observations.

\begin{figure*}
\includegraphics[width=\textwidth]{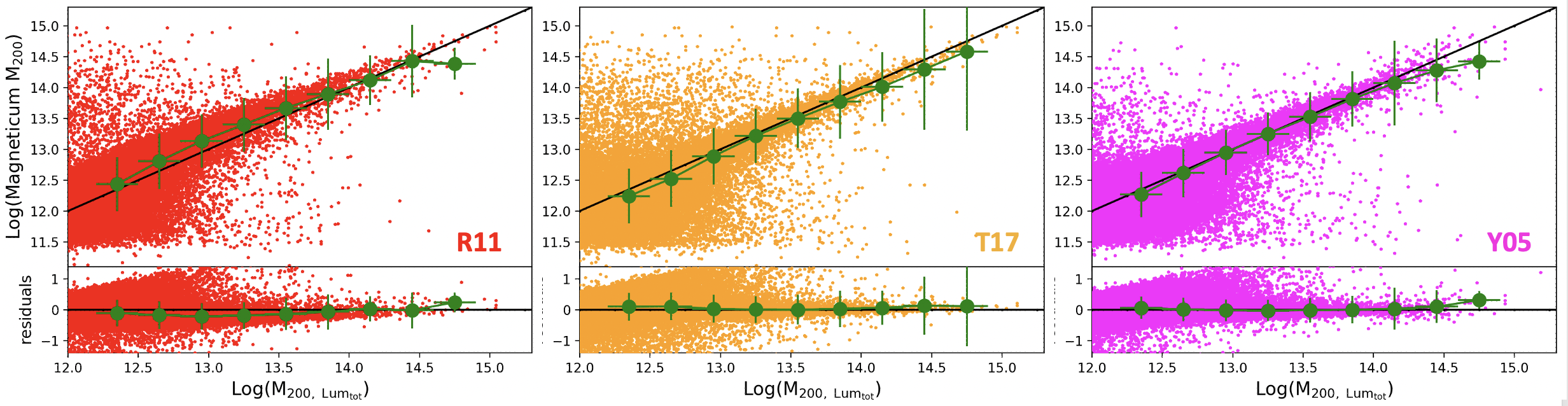}
\caption{The top panels show the comparison between the input halo mass $M_{200}$ and the group mass proxy based on the total luminosity scaled through the scaling relation of \cite{Popesso2005} for the three optically selected galaxy group samples: R11 (left panel), T17 (central panel) and Y05 (right panel. The solid line indicates the 1:1 relation. The green symbols indicate in all panels the mean $M_{200}$ obtained in bins of 0.3 dex width of $M_{lum}$, whilst the error bars indicate the dispersion. The bottom panel shows the residuals $\Delta=log(M_{lum})-log(M_{200})$ of the individual points and the mean values (green symbols) for the three group samples.}
\label{fig1}
\end{figure*}

\begin{figure*}
\includegraphics[width=\textwidth]{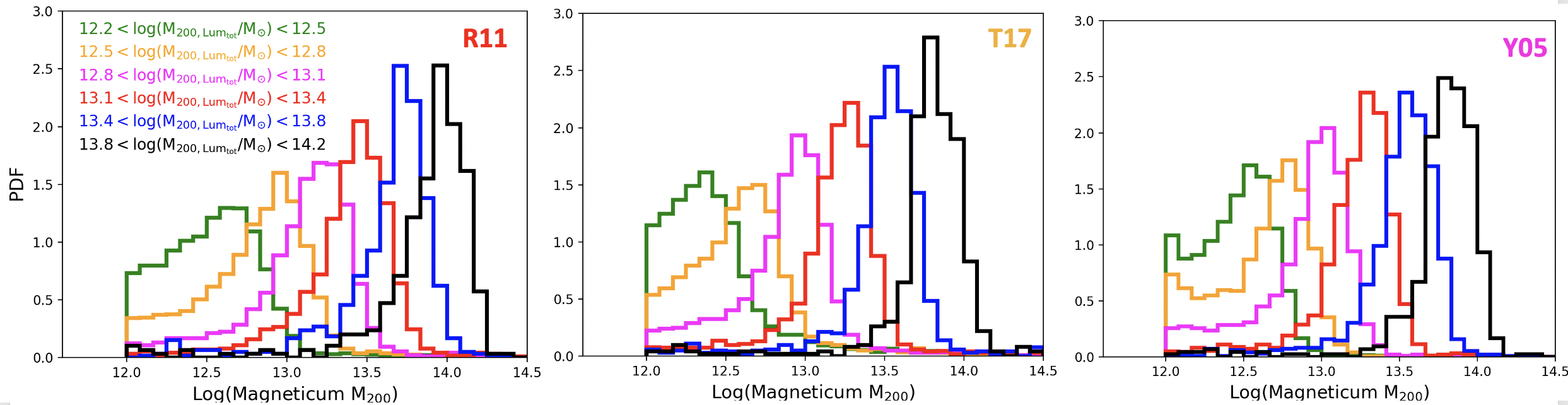}
\caption{The panels show the actual input halo mass $M_{200}$ distribution corresponding to different bins of halo mass proxy $M_{lum}$. The left panel shows the results of the R11 group sample, the central panel the T17 group sample, and the right panel the results of the Y05 sample. The color code is indicated in the left panel. }
\label{fig2}
\end{figure*}

\begin{table*}
\caption{Contamination of mass proxy for the R11, T17 and Y05 catalogs. For each catalog, we report in each halo mass proxy bin the corresponding values of the median of the input halo mass and the percentage levels of contamination by halos at masses higher and lower, receptively, of the bin limits. }             
\label{table1}      
\centering          
\begin{tabular}{c | c c c | c c c | c c c }     
\hline\hline       
$\Delta{Log{M_{200}}}$ & \multicolumn{3}{c}{R11} & \multicolumn{3}{c}{T17} & \multicolumn{3}{c}{Y05} \\ 
\hline                    
 & median & width & contamination & median & width & contamination & median & width & contamination \\
\hline
   $12.2-12.5$  & 12.43 & 0.43 & 32\%-34\% & 12.24 & 0.44 & 35\%-22\% & 12.24 & 0.37 & 47\%-27\%\\  
   $12.5-12.8$  & 12.70 & 0.45 & 27\%-40\% & 12.52 & 0.46 & 37\%-17\% & 12.60 & 0.4  & 41\%-25\%\\
   $12.8-13.1$  & 13.04 & 0.43 & 18\%-43\% & 12.89 & 0.46 & 26\%-16\% & 12.95 & 0.38 & 31\%-21\%\\
   $13.1-13.4$  & 13.32 & 0.43 & 16\%-43\% & 13.22 & 0.43 & 28\%-15\% & 13.26 & 0.36 & 23\%-18\%\\
   $13.4-13.7$  & 13.55 & 0.46 & 19\%-39\% & 13.52 & 0.46 & 24\%-14\% & 13.54 & 0.39 & 23\%-18\%\\
   $13.7-14.0$  & 13.83 & 0.45 & 18\%-34\% & 13.80 & 0.52 & 21\%10\%  & 13.81 & 0.43 & 24\%9\%\\
\hline                  
\end{tabular}
\end{table*}

\begin{figure*}
\includegraphics[width=\textwidth]{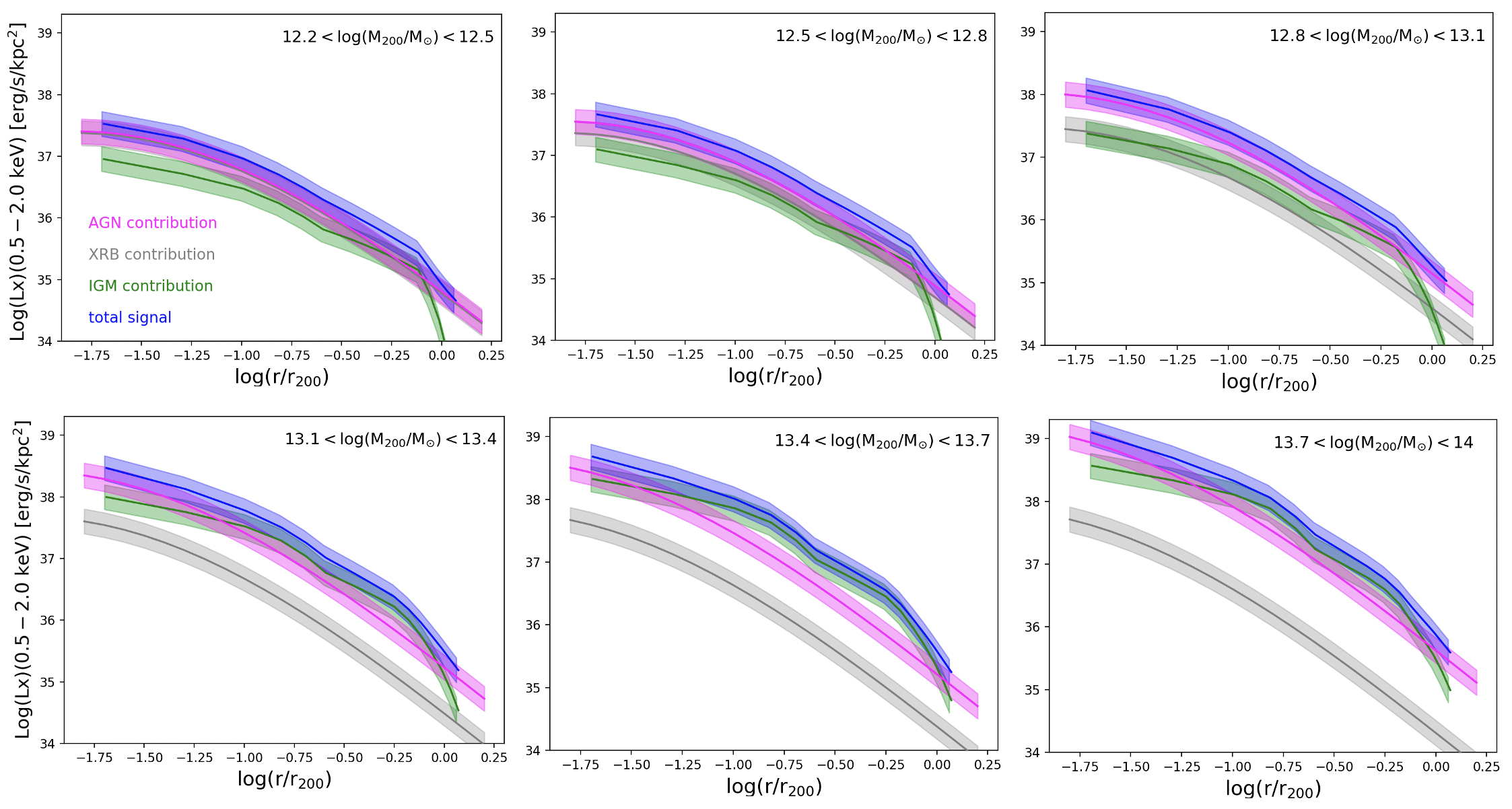}
\caption{Predicted Magneticum X-ray surface brightness profiles per bin of halo mass. The profiles are obtained by convolving the PHOX projected profiles with the eROSITA PSF. The blue-shaded regions indicate the total X-ray emission within 1$\sigma$ from the mean. The magenta blue-shaded regions indicate the AGN contribution, while the gray-shaded regions indicate XRB emission. The IGM emission is indicated by the green-shaded region. }
\label{phox}
\end{figure*}

\begin{table}
\caption{Contribution of emitting sources in halos per halo mass bin. The table shows the average contributions in all halos and in halos not containing eSASS detected point sources (PT). }             
\label{table2}      
\centering                          
\begin{tabular}{c | c c c | c c c}        
\hline\hline                 
\\    
$\Delta{Log{M_{200}}}$ & \multicolumn{3}{c}{All} & \multicolumn{3}{c}{no eSASS PT} \\ 
\hline                    
 & XRB & AGN & IGM & XRB & AGN & IGM  \\
\hline                        
   $12.2-12.5$  & 0.15 & 0.63 & 0.22 & 0.12 & 0.65 & 0.23 \\  
   $12.5-12.8$  & 0.11 & 0.64 & 0.25 & 0.09 & 0.64 & 0.26 \\
   $12.8-13.1$  & 0.05 & 0.66 & 0.30 & 0.04 & 0.67 & 0.29 \\
   $13.1-13.4$  & 0.03 & 0.39 & 0.58 & 0.02 & 0.37 & 0.61 \\
   $13.4-13.7$  & 0.01 & 0.25 & 0.74 & 0.01 & 0.21 & 0.78 \\
   $13.7-14.0$  & 0.01 & 0.26 & 0.73 & 0.01 & 0.21 & 0.78 \\
\hline                                   
\end{tabular}
\end{table}

\begin{figure}
\includegraphics[width=0.5\textwidth]{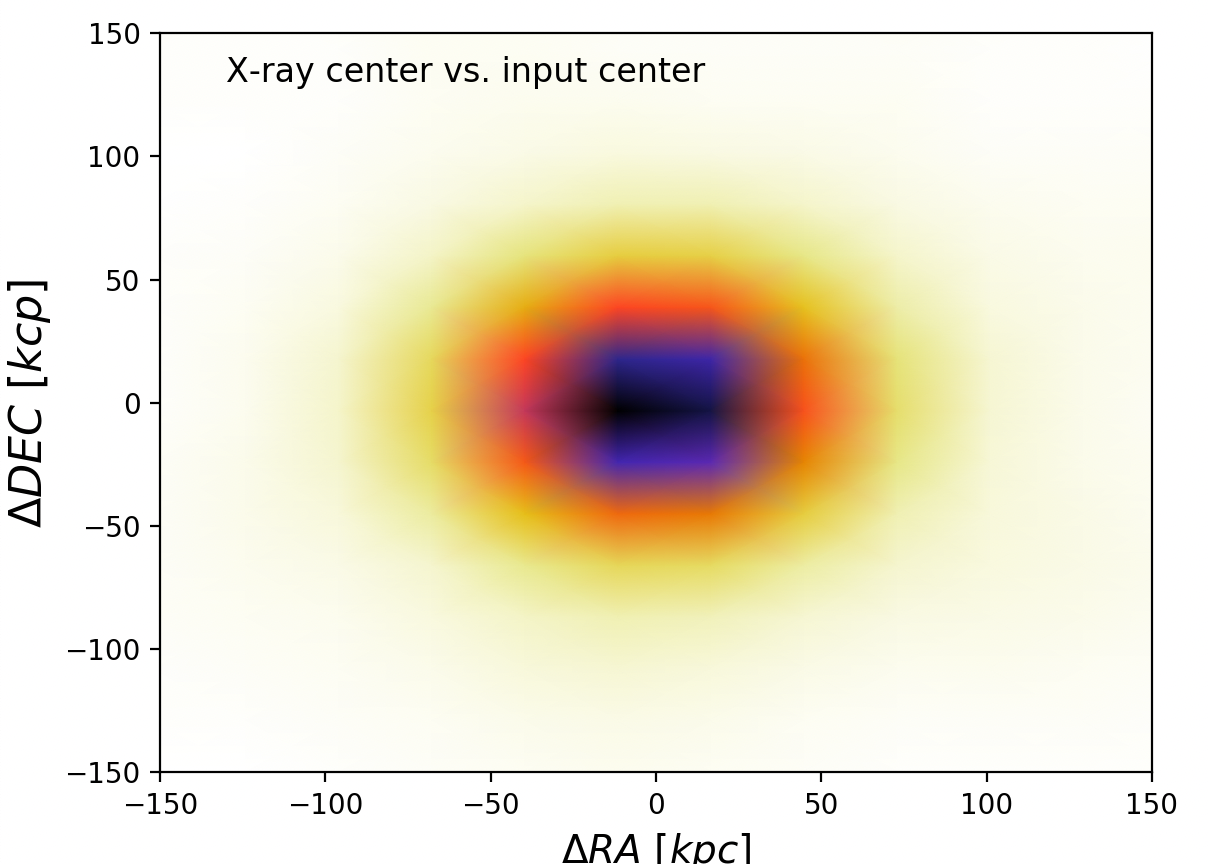}
\caption{The figure shows the difference in $\Delta{RA}$ and $\Delta{DEC}$, estimated in kpc, between the center of the eSASS detected extended emission and the center of the input halo.}
\label{fig3}
\end{figure}

\begin{figure*}
\includegraphics[width=\textwidth]{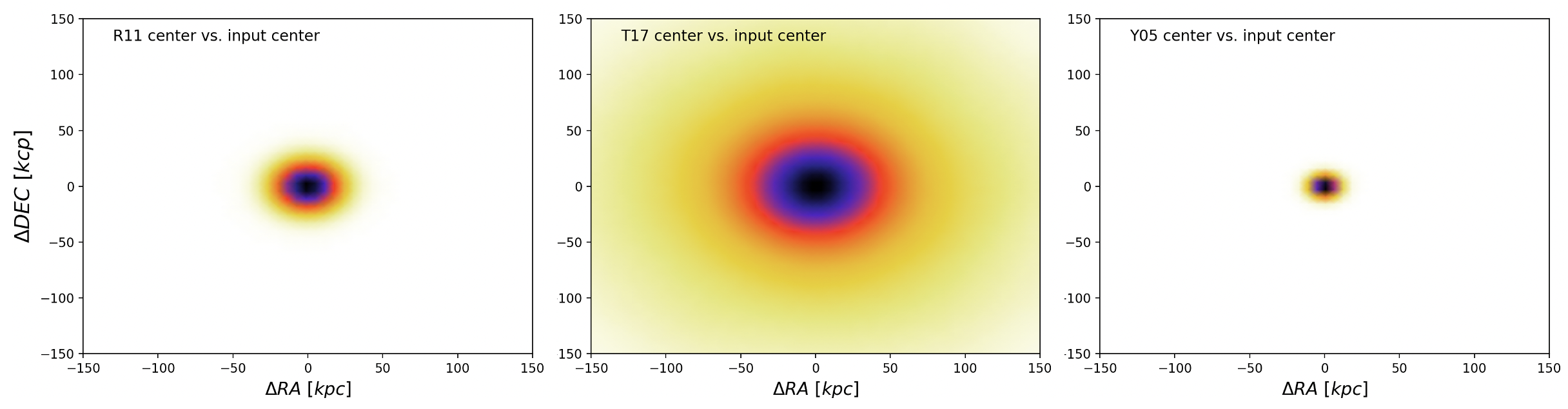}
\caption{The figure shows the difference in $\Delta{RA}$ and $\Delta{DEC}$, estimated in kpc, between the center of the detected optical group and the center of the input halo (left panel), the coordinates of the eSASS X-ray detection and the center of the input halo (central panel), and the center of the detected optical group and of the eSASS X-ray detection.}
\label{fig4}
\end{figure*}

\begin{figure*}
\includegraphics[width=\textwidth]{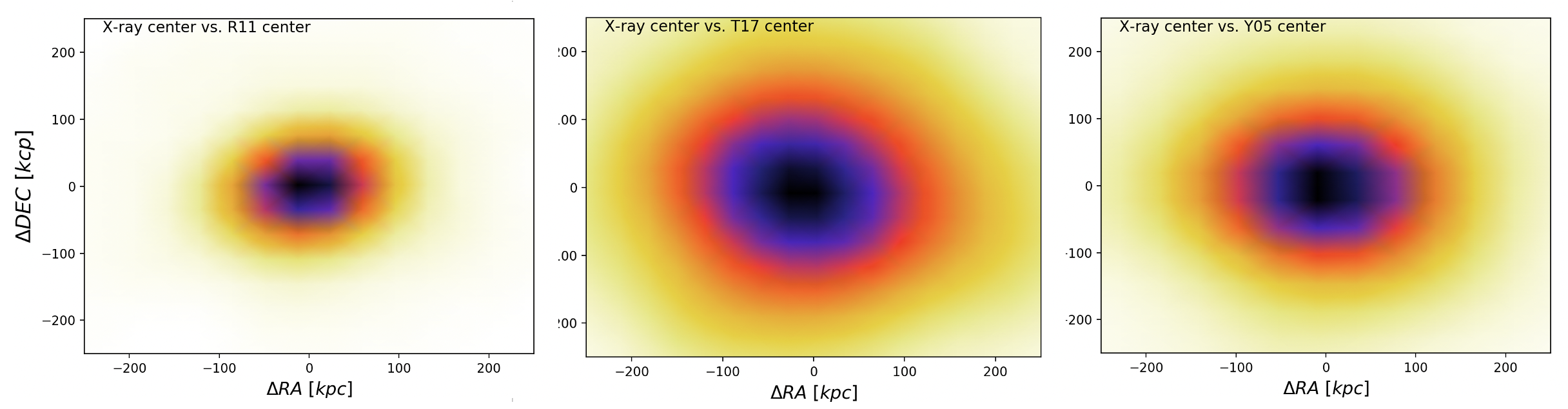}
\caption{The figure shows the difference in $\Delta{RA}$ and $\Delta{DEC}$, estimated in kpc, between the center of the detected optical group and the center of the input halo (left panel), the coordinates of the eSASS X-ray detection and the center of the input halo (central panel), and the center of the detected optical group and of the eSASS X-ray detection.}
\label{fig5}
\end{figure*}

\subsubsection{Completeness and contamination}

Figure \ref{fig1} shows the completeness and contamination of the sample compared to the input halo catalog. Completeness is estimated as $N_{Matches}(M_{200})/N_{halos}(M_{200})$, where $M_{200}$ is the Magneticum input halo mass, $N_{Matches}(M_{200})$ is the number of detected groups matched in coordinates, redshift, and mass to the input halo catalog, and $N_{halos}(M_{200})$ is the number of halos in the input catalog at $M_{200}$. Contamination is estimated as $N_{unmatched}(M_{obs})/N_{group}(M_{obs})$, where $M_{obs}$ is the observed mass proxy of the group sample based on the group optical luminosity, $N_{unmatched}(M_{obs})$ is the number of groups without a match in the input halo catalog, and $N_{group}(M_{obs})$ is the total number of groups detected at mass $M_{obs}$. As highlighted in \citet{Marini24b}, the completeness of the R11 group sample remains above 90\% down to $log(M_{200}) \sim 10^{13.5} M_{\odot}$, decreases to 80\% at $log(M_{200}) \sim 10^{13} M_{\odot}$, and drops to 50\% at $log(M_{200}) \sim 10^{12} M_{\odot}$. The T17 sample exhibits a very similar behaviour, while the Y05 sample remains highly complete ($>90\%$) down to $10^{12}$ $M_{\odot}$. The contamination due to spurious detections remains below 10\% down over the entire mass range for all group samples.

These results are qualitatively consistent with those of \cite{Robotham2011}. Quantitatively, the levels of completeness and contamination presented here are higher and lower, respectively, than in \cite{Robotham2011}, due to the lower redshift cut ($z < 0.2$) of the mock galaxy sample compared to GAMA, which extends to $z \sim 0.5$.

For comparison, we show also the completeness of the X-ray {\it{eSASS}} selection applied to the mock eROSITA observations as described in \citet{Marini2024a}. In this case, the completeness drops quickly to less than 50\% below $10^{14}$ $M_{\odot}$ and below 10\% at $10^{13.5}$ $M_{\odot}$, whilst the contamination increases to 20\% in the same mass range. 

\subsubsection{The optical halo mass proxy}

According to \citet{Marini24b}, the best proxy for the input $M_{200}$ is the total luminosity, which is converted into a mass proxy, $M_{lum}$, using the scaling relation of \cite{Popesso2005}, when estimated within the same absolute magnitude range. While total stellar mass is also a good proxy, it has a slightly larger scatter. The mass derived from velocity dispersion is unreliable for groups with fewer than 10 galaxy members \citep[see][for a more detailed analysis]{Marini24b}.

The scatter of the $M_{lum}-M_{200}$ relation varies with mass, being relatively small ($\sim$0.14 dex) for $M_{lum}>10^{13.5}$ $M_{\odot}$ and increasing up to 0.3 dex at lower values of the proxy (Fig. \ref{fig1}). This implies that when stacking the optically selected groups in $M_{lum}$ bins, there is contamination by both lower and higher mass groups. For a fixed bin width of 0.3 dex in $M_{lum}$, this contamination increases steadily from $M_{lum}>10^{14}$ $M_{\odot}$ to $M_{lum}\sim10^{12.5}$ $M_{\odot}$ for all optically selected group samples. The contamination is not symmetrical as shown in Fig. \ref{fig2}. The net effect, as reported in Table \ref{table1}, is that a bin width of 0.3 dex in $\Delta{M_{lum}}$ corresponds to an actual bin width on average of $\sim$0.45 for R11 and T17 and $\sim$0.38 dex in $\Delta{M_{200}}$ for Y05. The actual bin width is estimated as the width capturing 90\% of the systems entering the $\Delta{M_{lum}}$ bin. The median mass of the bin is slightly underestimated for R11. This leads to a larger contamination by more massive systems, while the opposite, a slightly larger contamination by less massive systems, is observed for T17 and Y05. This will be considered when plotting the error bars of the mean $M_{lum}$ per bin in further analyses. 


We also note that the largest contamination observed at MW-sized halos ($M_{lum}\sim10^{12.5}$ $M_{\odot}$) is mainly due to the lack of any calibration between proxies and $M_{200}$ at this mass scale. Indeed, all available calibrations extend to relatively massive groups with masses larger than $10^{13}$ $M_{\odot}$.

\subsection{The synthetic eROSITA observations}
The photon list of all X-ray emitting components, including hot gas, AGN, and X-ray binaries, is generated using \texttt{PHOX} \citep{biffi_observing_2012, biffi_investigating_2013,VZS23} for the same light cone L30 used to create the mock galaxy catalog. \texttt{PHOX} computes X-ray spectral emission based on the physical properties of the gas, black hole (BH), and stellar particles in the simulation. The photon list is created for all components up to a redshift $z < 0.2$, providing the X-ray counterpart of the galaxy mock catalog in the local Universe. Detailed modeling of the components can be found in \citet{Marini2024a}.

The synthetic photon lists are used as input files for the Simulation of X-ray Telescopes (\texttt{SIXTE}) software package \citep[v2.7.2;]{dauser_sixte_2019}. \texttt{SIXTE} incorporates all instrumental effects, including the PSF, redistribution matrix file (RMF), and auxiliary response file (ARF) of the instrument \citep{Predehl2021}. It can also model eROSITA's unvignetted background component due to high-energy particles \citep{LiuAng2022}. We perform mock observations of eRASS:4 in scanning mode using the theoretical attitude file for the three components separately, then combine the event files. In LC30, the simulated background in all seven TMs is based on \cite{LiuTeng2022} and represents the spectral emission from all unresolved sources, rescaled to the eRASS:4 depth in line with the simulated emission of the individual X-ray emitting components.

The simulated eROSITA data is processed through eSASS as described in \cite{merloni_srgerosita_2024}. The event files from all emission components and eROSITA TMs are merged and filtered for photon energies within the $0.2-2.3$ keV band. The filtered events are binned into images with a pixel size of $4\arcsec$ and $3240 \times 3240$ pixels. These images correspond to overlapping sky tiles of size $3.6 \times 3.6$ deg$^{2}$, with a unique area of $3.0 \times 3.0$ deg$^{2}$. The LC30 is covered by 122 standard eRASS sky tiles. A detailed description of the data reduction is provided in \citet{Marini2024a}, along with the corresponding X-ray catalog of extended and point sources (PT) provided by \texttt{eSASS}.

\citet{Marini24b} thoroughly analyze the completeness and contamination of the extended emission catalog, finding that, after matching the X-ray detections with the input halo catalog, the sample's completeness drops below 80\% at $M_{200} \sim 10^{14} M_{\odot}$, reaches 45\% at $\sim 10^{13.5} M_{\odot}$, and no sources are detected at $\sim 10^{13} M_{\odot}$ (see Fig. \ref{fig0}). The contamination is negligible in the cluster mass range and is about 20\% for $10^{13} M_{\odot} < M_{200} < 10^{14} M_{\odot}$.

\begin{figure}
\includegraphics[width=0.5\textwidth]{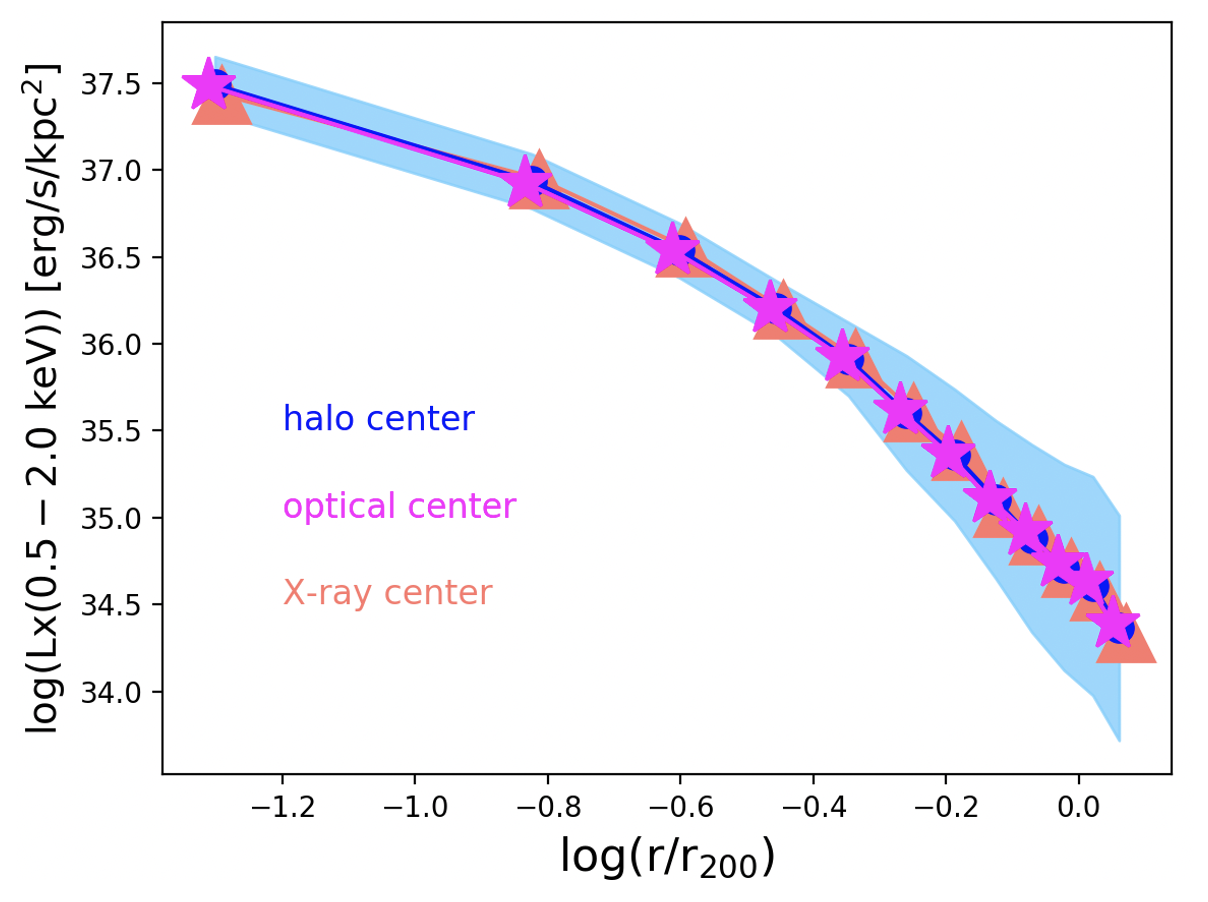}
\caption{The figure shows the comparison of the average projected PHOX profiles convolved with the eROSITA PSF and centered on the input halo center (blue points), the center coordinates of the R11 mock optically selected group sample (magenta symbols), and the X-ray center (orange triangle). The shaded region indicates the overlapping 1$\sigma$ region.}
\label{fig6}
\end{figure}

\section{The predicted eROSITA X-ray surface brightness profile}
In this section, we provide the predicted emission associated with halos generated from PHOX on the L30 light cones. These predictions serve not only as benchmarks for comparison with observations but also as a basis to test the results of the stacking analysis, which underpins the observational findings.

We generate the average X-ray surface brightness profiles of all components associated with halos, including IGM, AGN, and X-ray binaries, in several halo mass bins of 0.3 dex width and convolved with the eROSITA PSF, with $HEW=30$ arcsec, averaged in the 0.2-2.3 keV band \citep{merloni_srgerosita_2024}. The individual profiles are created in annuli of $r/r_{200}$ and averaged azimuthally. Fig. \ref{phox} shows the Magneticum prediction in six halo mass bins covering the entire galaxy group mass range, from MW-type galaxy groups to poor clusters with halo masses of $10^{14}$ $M_{\odot}$. Table \ref{table2} indicates the fractional contribution of each X-ray emitting component as a function of the halo mass bin.

As a caveat, we note that in this analysis, we include systems in the lowest halo mass bins, with halo masses in the ranges $10^{12.2}-10^{12.5}$ $M_{\odot}$ and $10^{12.5}-10^{12.8}$ $M_{\odot}$. These are representative of MW-sized halos, corresponding to galaxies with stellar masses in the range $10^{10.5}-10^{11}$ $M_{\odot}$ \citep{Berhoozi2019}. However, it is important to mention that the virial temperature of the gas in such low-mass halos is below the effective area of the soft band of eROSITA. Consequently, the X-ray surface brightness profiles provided here in the $0.5-2$ keV eROSITA soft band reflect the gas at the high-temperature tail of the CGM or IGM temperature distribution. Nonetheless, we provide this calibration due to recent eROSITA stacking results for this halo and stellar mass range \citep{Comparat2022, Zhang24a}.

As visible in Fig. \ref{phox} and quantified in Table \ref{table2}, the IGM and AGN emissions remain the main components at all halo masses. The sum of their contributions accounts for nearly the totality of the group X-ray emission. The contribution from X-ray binaries remains very low due to the low SFR values of nearby galaxies and the dominance of quenched galaxies in the local Universe. This contribution is somewhat significant (10-15\%) below $M_{200}\sim10^{13}$ $M_{\odot}$ and becomes negligible at higher halo masses. The AGN component is the dominant component for all groups with masses below $10^{13}$ $M_{\odot}$, nearly equals the IGM contribution for groups at $M_{200}\sim10^{13}$ $M_{\odot}$, and becomes less significant at larger halo masses. In all cases, the contribution is primarily due to the AGN activity of the central galaxies, while the sporadic AGN activity of satellites is averaged out azimuthally.

\section{Reliability of the stacking procedure}
As shown in Fig. \ref{fig0}, the eSASS detections account for only a marginal fraction of the galaxy group population in the local Universe. From an observational perspective, accessing the average properties of the gas in groups requires stacking galaxy groups selected using different techniques. In this section, we test the results of stacking optically selected groups in the eROSITA data.

The result of any stacking procedure and its interpretation highly depends on how well the prior sample is known and characterized in terms of selection effects. Additionally, one should take into account uncertainties due to center coordinates, assumptions of gas temperature and metallicity, the halo mass proxy, and contamination by other X-ray emitting sources such as AGN and XRB.

In this case, the prior catalogs provided by the three different optically selected samples have been thoroughly analyzed in terms of completeness and contamination in \citet{Marini2024a}. The high level of completeness and low contamination allow capturing the bulk of the halo population at each halo mass through stacking. Here, we analyze the effect of using optical centers, the impact of assuming an average temperature and metallicity on the gas emissivity, the contamination from other X-ray emitting sources, and the uncertainty of the halo mass proxy in the stacking procedure outlined in \cite{Popesso24}.

\begin{figure}
\includegraphics[width=0.5\textwidth]{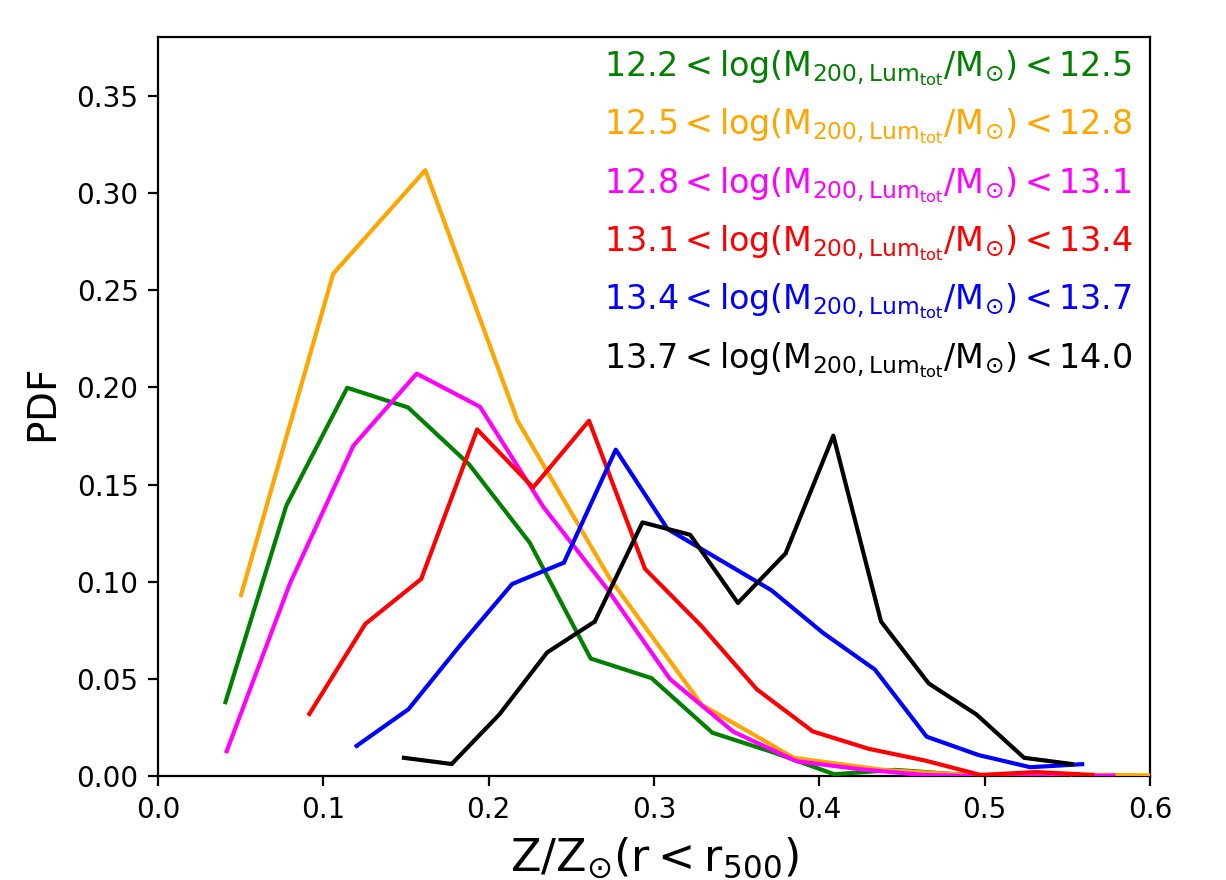}
\caption{The figure shows the distribution of the mass-weighted metallicity of the Magneticum halos estimated within $r_{500}$ in different halo mass bins. The histograms are color-coded as described in the figure according to the halo mass bin.}
\label{metallicity}
\end{figure}

Briefly, the stacking is done by averaging the background subtracted surface brightness profiles within the same annuli around the group center. The background is measured in a region between 2 to 3 Mpc from the group center. All events flagged as point source in each annulus are excluded and the corresponding annulus area is corrected for the excluded point source area. All groups containing a point source or contaminated by close neighbours within $2\times r_{200}$ are excluded from the prior sample for the stacking. The X-ray luminosity from the stacked signal is derived in the 0.5-2 keV band by selecting only events with a rest frame energy in the selected band at the median redshift of the prior sample. We refer to \cite{Popesso24} for a more detailed description of the procedure.

\begin{figure*}
\includegraphics[width=\textwidth]{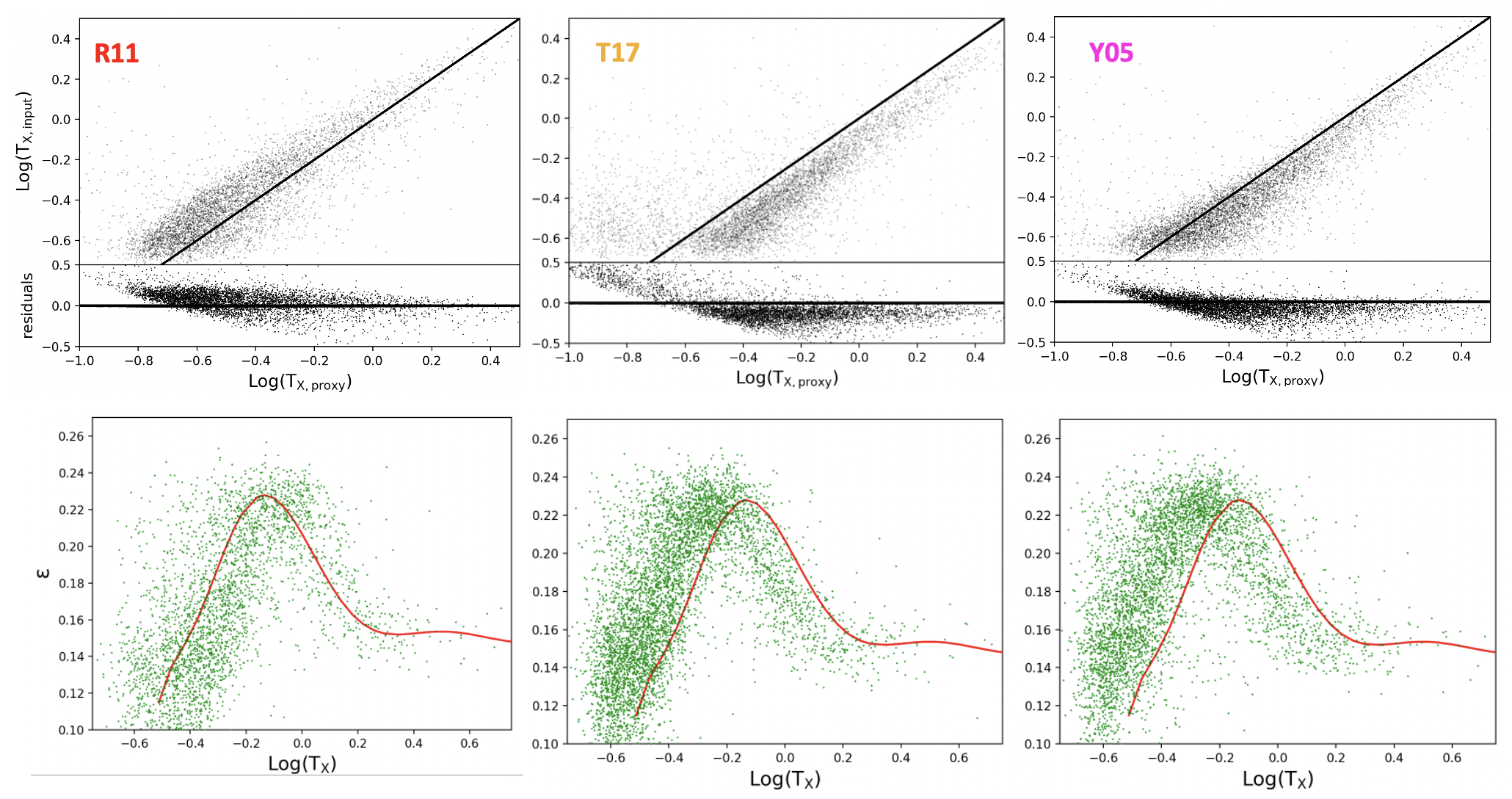}
\caption{{\it{Upper panels}}: Each panel shows the relation between the input mass-weighted gas temperature derived within $r_{500}$ for the Magneticum halos and the temperature derived from the optical halo mass proxy through the scaling relation of \cite{Lovisari2015} for each prior catalog: R11 (left panel), T17 (central panel), and Y05 (right panel). Each panel also shows the residuals in dex as a function of the gas temperature proxy. {\it{Bottom panels}}: The panels show the expected temperature-emissivity relation, at fixed metallicity (0.3$Z_{\odot}$), estimated at the input mass-weighted gas temperature derived within $r_{500}$ for the Magneticum halos and the X-ray emissivity (red curve in each panel). The green points indicate the emissivity corresponding to the temperature proxy obtained from the optical halo mass proxy of the different optically selected group samples (R11 left panel, T17 central panel, and Y05 right panel) and estimated for the same metallicity.}
\label{emissivity}
\end{figure*}

\subsection{Effect of X-ray-optical center residuals in the stacking procedure}

Figure \ref{fig3} shows the distribution of the residuals $\Delta{RA}$ and $\Delta{Dec}$ in kpc between the X-ray center of the eSASS detections and the center of the matched halo in the input catalog. The scatter remains under 10 kpc, which is significantly smaller than the corresponding value in kpc of the FWHM of the eROSITA PSF, approximately 25 arcsec, at all redshifts. These results are consistent with \cite{Seppi2022}.

Figure \ref{fig4} shows the residuals in the center coordinates between the optically detected group center and the input halo center for the three galaxy group samples. For the majority of the R11 and Y05 groups, the center coincides with the position of the brightest group galaxy (BGG) almost at rest. \citet{Marini24b} find that all algorithms are highly efficient in identifying the BGG, with a success rate above 90\% at nearly all halo masses. Due to this high efficiency, the detected optical center and the input halo coordinates coincide, with $\Delta{RA}$ and $\Delta{Dec}$ close to zero. The standard deviation around the 0 residuals is approximately 15 kpc for R11, 45 kpc for T17, and 7 kpc for Y05. For the remaining 10\% of cases, the residuals are below 50 kpc.

Figure \ref{fig5} shows the residuals between the X-ray and the optical center. Interestingly, the scatter of the residuals between the R11 optical center and the X-ray center is very small (less than 15 kpc). This may suggest a correlation in the deviation of the optical and X-ray centers relative to the input halo center. Although the Y05 algorithm is the most efficient in retrieving the input center, the scatter in $\Delta{RA}$ and $\Delta{Dec}$ is around 50 kpc. The largest discrepancy is observed for the T17 sample, with an average residual of approximately 78 kpc.

Figure \ref{fig6} illustrates the comparison between the average surface brightness profile derived by averaging the input X-ray surface brightness profiles of PHOX, convolved with the eROSITA PSF by using different centers: the input halo, and the X-ray and the optical centers. This example considers all groups simultaneously identified in the X-rays by eSASS and in the optical by the R11 algorithm, within a halo mass bin of $10^{13.5} M_{\odot} < M_{200} < 10^{13.8} M_{\odot}$. The profiles exhibit remarkable consistency. Very similar results are obtained also for the T17 and Y05 algorithms. The result demonstrates that the effect of the residuals in averaging the X-ray surface brightness profile is washed out by the large eROSITA PSF. 

Since we have not observed any effect on the input PHOX profiles, this ensures that no effects will be propagated into the stacking analysis of the mock eROSITA observations. 

\begin{figure*}
\includegraphics[width=\textwidth]{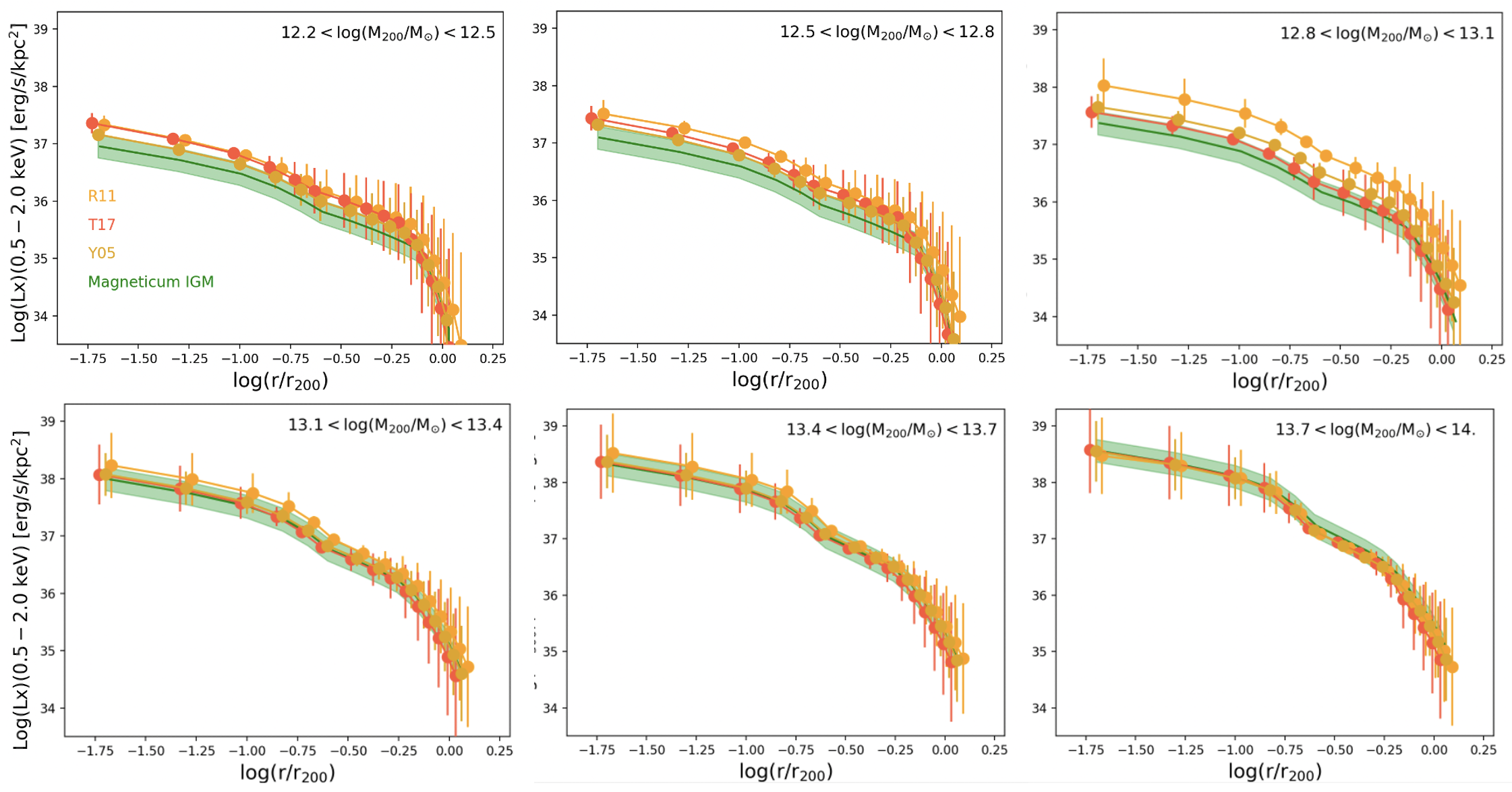}
\caption{The figure shows the comparison in different halo mass bins between the PHOX input IGM profiles convolved with the eROSITA PSF and averaged in input halo mass bins (green shaded region), with profiles averaged in halo mass proxy bins (colored symbols). The green-shaded region indicates the 1$\sigma$ error. The colored points show the mean PHOX profiles obtained from the different prior catalogs (yellow for R11, orange for T17, and gold for Y05). The error bars indicate the $1\sigma$ uncertainty.}
\label{stacking}
\end{figure*}

\subsection{AGN and XRB correction to isolate the IGM emission} 
As outlined in \cite{Popesso24}, all groups with a detected eSASS point source within twice $r_{200}$ are discarded from the prior samples. This is done not only to mitigate the AGN contribution and isolate the IGM component but also to clean the area around the group to properly measure and remove the background.

Using the input PHOX information, we measure the average contribution of the individual X-ray components in the mean group profile by discarding all halos with an eSASS-detected point source within $2\times r_{200}$ from the sample. Table \ref{table2} reports the contribution of the X-ray components within $r_{200}$ after removing all halos containing a PT within $2\times r_{200}$. We note that there is a negligible change in AGN contamination after PT removal. This indicates that the contamination is not due to the few relatively bright AGN detected by eSASS in our eRASS:4-like mock observations. Instead, AGN fainter than the eRASS:4 detection threshold are responsible for the totality of the contamination.

We observe that discarding groups with a PT within $2\times r_{200}$ leads to lowering the noise of the background measurement. Indeed, removing PT enables us to improve the accuracy of the background estimate by 35\%. We conclude that removing halos contaminated by PT detections has a marginal effect on mitigating AGN contamination but is essential to increase the accuracy of the background estimate.

Nevertheless, correcting for AGN and XRB contamination is essential to properly isolate the IGM contribution. This can only be done by modeling the halo occupation distribution and the conditional X-ray luminosity function of the AGN population. 
Many models are available for the halo occupation distribution of AGN at different redshifts \citep[e.g.,][]{Georgakakis2019,krumpe2018,Aird2021,powel2022,krumpe2023,comparat2023}. These models can reproduce the clustering properties and the X-ray luminosity function of the observed AGN population fairly well. However, all models are constrained either by observations at higher redshifts \citep[e.g.,][]{Georgakakis2019,Aird2021,comparat2023} or by relatively bright local AGN \citep[e.g.,][]{krumpe2018,powel2022}. Thus, they require extrapolation to lower redshift, as in our case ($z < 0.2$), or to lower luminosities. Since there is no easy way to discern which model performs better in this parameter range, we rely on the Magneticum results. Indeed, \cite{hirschmann_cosmological_2014} show that the AGN model implemented in Magneticum can broadly match observed black hole properties of the local Universe, including the observed soft and hard X-ray luminosity functions of AGN and clustering properties. \cite{Marini2024a} also shows that the analysis of the eRASS:4-like mock observations based on the Magneticum light cones reproduces well the soft X-ray AGN luminosity function of \cite{marchesi_mock_2020}. We thus correct the X-ray surface brightness profile obtained through the stacking analysis by subtracting a PSF with a signal equal to the percentage of X-ray luminosity due to AGN contamination per halo mass bin, as indicated in Table \ref{table2}.

To remove the residual XRB contamination, we apply the following approach. The XRB contribution, in nearly all cases, is due to the mean SFR of the central galaxies. Satellite galaxies in the local Universe tend to be quiescent or highly quenched, occupying regions well below the Main Sequence (MS) of star-forming galaxies \citep{popesso2019}. Thus, their SFR activity contributes very little to the XRB emission, becoming negligible when averaged azimuthally to obtain the surface brightness profile. Central galaxies in MW-sized halos, on the other hand, tend to be MS galaxies, as shown in \cite{popesso2019}. Therefore, since in most cases, the optical center coincides with the central galaxy location, the XRB contribution due to the star formation activity of the central galaxy can be modeled by a PSF rescaled to reproduce the X-ray luminosity of this contribution.

To estimate this, we use the scaling relation from \cite{lehmer2016}, which links the X-ray luminosity in the 0.5-2 keV band to the galaxy SFR. We use the mean SFR of the central galaxies in each halo mass bin to estimate the corresponding mean X-ray luminosity. We then subtract a PSF rescaled to this luminosity from the stacked profile to remove the IGM contribution. This approach nicely reproduces the same level of XRB contamination estimated through PHOX in each halo mass bin.

\begin{figure*}
\includegraphics[width=\textwidth]{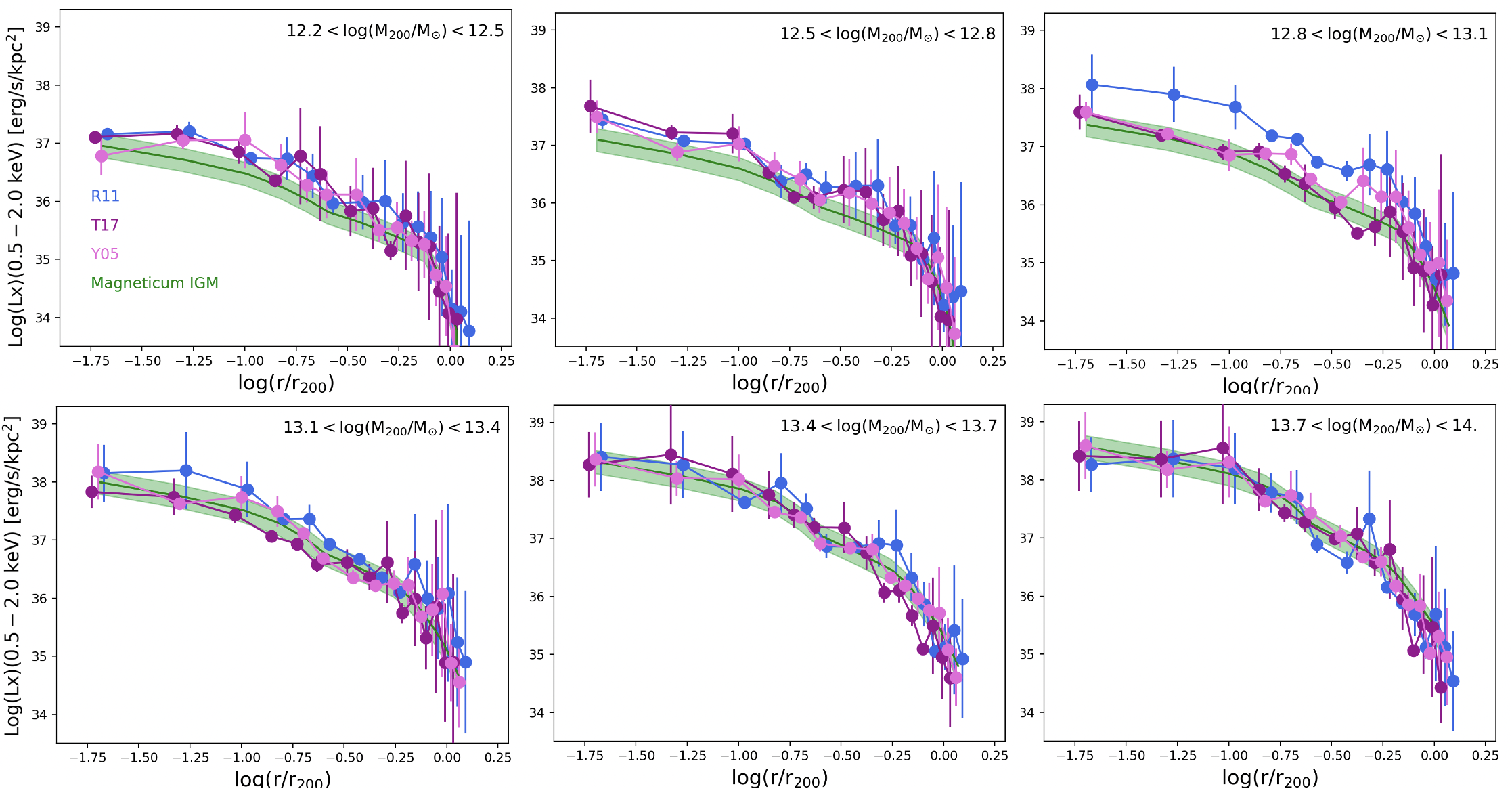}
\caption{The figure shows the comparison in different halo mass bins between the PHOX input IGM profiles convolved with the eROSITA PSF and the results of the stacking procedure on the mock observations. The PHOX profiles are averaged in bins of input halo mass, while the stacked profiles are obtained in bins of halo mass proxy of the related prior catalog (blue for R11, violet for T17, and pink for Y05).}
\label{after_stacking}
\end{figure*}

%

\subsection{The group X-ray emissivity}
\label{emissivity_1}

The X-ray emissivity is the energy emitted per time and volume by the hot gas in galaxy systems, typically described as an optically thin plasma in collisional ionization equilibrium. This is a key element when deriving the gas mass profile from the X-ray surface brightness profile of halos. The emissivity is defined as $n_p n_e \Lambda(T, Z)$, where $n_p$ and $n_e$ are the proton and electron densities, respectively, and $\Lambda(T, Z)$ is the gas cooling function, which depends on the emission process, gas temperature, and metallicity. As shown in \cite{2021Univ....7..254L}, at temperatures higher than approximately 3 keV, the main emission mechanism is thermal bremsstrahlung. At this scale, the emissivity in the 0.5-2 keV band considered here is nearly independent of temperature and metallicity. However, at lower temperatures, within the group regime, line cooling becomes significant, and the emissivity strongly varies as a function of metallicity and temperature.

The temperature and metallicity of the hot gas in galaxy groups can only be measured by modeling the gas X-ray spectrum and its emission lines. However, this information is not available for the undetected systems in the optically selected prior catalog. Thus, to estimate the emissivity and derive the hot gas mass profile from the observed X-ray surface brightness profile, we must assume a mean temperature and metallicity. While an average temperature for groups can be derived from existing scaling relations \citep[see][]{Lovisari2015, 2021Univ....7..254L}, very little is known about the global distribution and average gas metallicity, as well as the shape of possible spatial gradients in low-mass systems \citep[see also][]{Sun2009, Mernier2017, lovisari19}.

In this analysis, we assume a mean temperature derived from the group halo mass proxy using the \cite{Lovisari2015} $M-T_x$ scaling relation, similar to how we would handle real observations. This assumption is justified for our eROSITA mock observations because Magneticum simulations accurately reproduce the \cite{Lovisari2015} $M-T_x$ scaling relation, as shown in \cite{Marini2024a}. We assume a mean metallicity of $0.3Z_{\odot}$ for all groups and neglect any temperature and metallicity profiles within the groups, although they are present in the Magneticum X-ray groups. As shown in Fig. \ref{metallicity}, the assumption of $0.3Z_{\odot}$ is reasonable for the Magneticum simulation. The figures shows the distribution of mass-weighted metallicity estimated from the hot gas particles in the simulation within $r_{500}$ per halo mass bin. This slightly varies from 0.17$Z_{\odot}$ at approximately $3 \times 10^{12}$ $M_{\odot}$ to 0.37$Z_{\odot}$ at approximately $10^{14}$ $M_{\odot}$, but the trend is poorly significant. Additionally, we point out that the gas metallicities at the group mass scale treated here, from MW-sized halos to massive groups, are very poorly constrained and measured only in X-ray-selected groups, which might provide a biased picture.

The upper panels of Fig. \ref{emissivity} compare the input mass-weighted temperature measured from the hot gas particles in the simulation within $r_{500}$ with the temperature derived from the best halo mass proxy based on the total luminosity of each prior catalog. The temperature proxy agrees well with the input temperature for the R11 catalog.  The T17 and Y05 catalogs tend to largely underestimate the temperature below to 0.25 keV and overestimate above this threshold by 20-30\%. This aligns with the slight underestimation of the halo mass proxy compared to the input halo mass for the T17 and Y05 catalogs, as shown in Fig. \ref{fig1}. The lower panel of the same figure shows how this effect translates into an error in the X-ray emissivity. The red curve in the plot shows the emissivity estimated in the 0.5-2 keV band, assuming a fixed metallicity of $0.3Z_{\odot}$ as in \cite{asplund09}. The green points indicate, for each prior catalog, the effect on the emissivity at the same metallicity and energy band based on the temperature mass proxy. For the R11 catalog, the use of the temperature proxy leads to a percentage uncertainty in the X-ray emissivity of 23\% without evident systematics. For the T17 and Y05 catalogs, the net effect is an overestimation of 20\% and 0.27\% of the emissivity below 0.6 keV and an underestimation of approximately 10\% between 0.8 and 1.6 keV. Nearly the same values are found at higher metallicity, while at lower values, the effect is marginal in all cases.

For the sake of this experiment, such variation in metallicity has only a minor effect on the mean emissivity per halo mass bin. Indeed, the effect of the temperature proxy uncertainty is larger. Thus, the assumption of a $0.3Z_{\odot}$ metallicity is reliable. However, we point out that a larger spread might be present in the observed groups.

\begin{figure*}
\includegraphics[width=\textwidth]{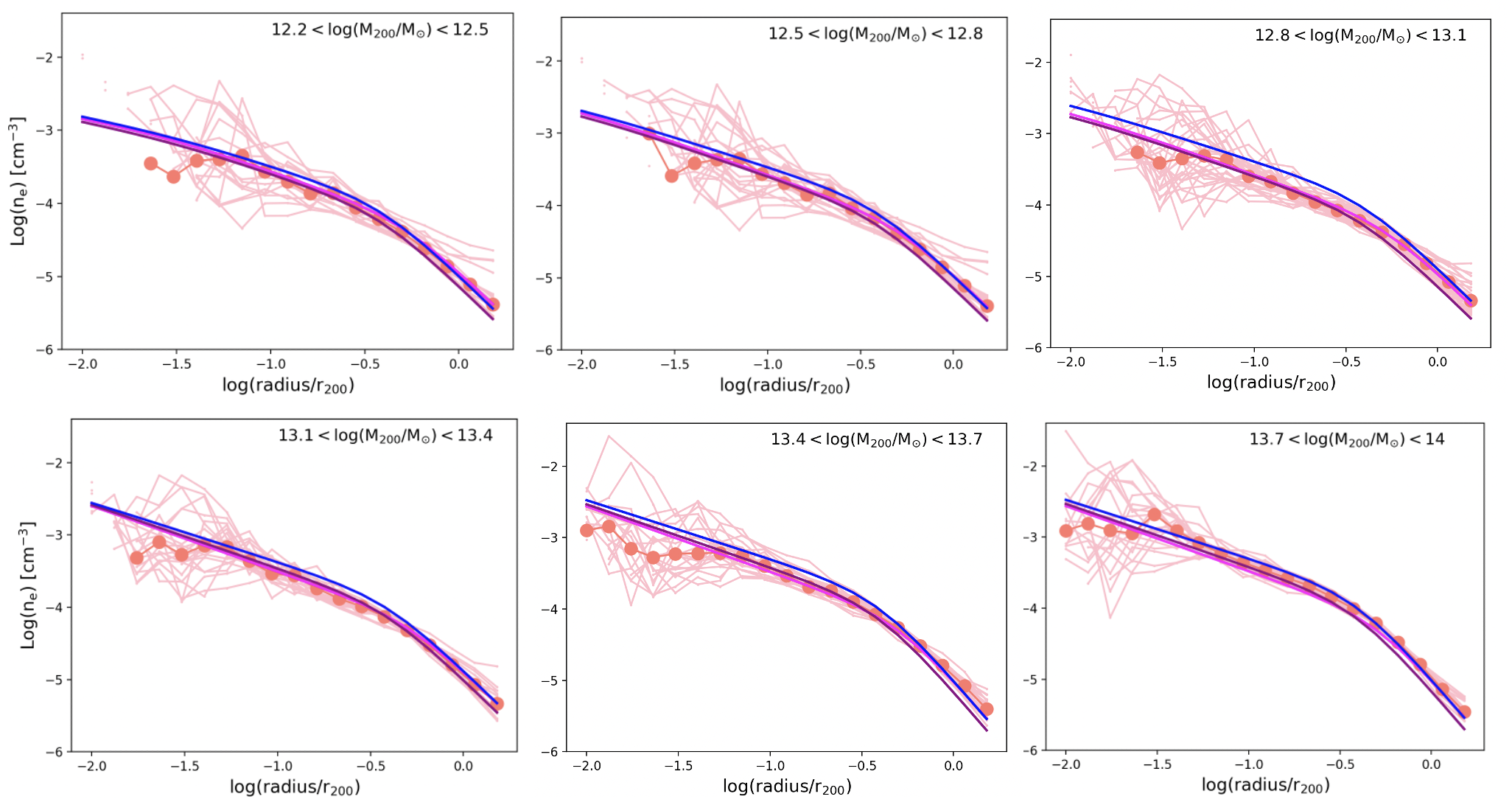}
\caption{The figure shows the comparison in different halo mass bins between the electron density profiles predicted by Magneticum (light pink points and solid curves) and the best fit profiles obtained from the stacked X-ray surface brightness profiles of Fig. \ref{after_stacking} of the three optical group catalogs of R11 (blue curve), Y05 (magenta curve) and T17 (purple curve).}
\label{ne_profile}
\end{figure*}

\subsection{The uncertainty of the halo mass proxy}

To test the effect of uncertainty in the halo mass proxy, we measured the average PHOX profiles convolved with the eROSITA PSF, binned by halo mass based on the halo mass proxy of different samples.

The retrieved IGM contribution is shown for each halo mass proxy bin and prior catalog in Fig. \ref{stacking} and compared with the predicted IGM profile from Magneticum as shown in Fig. \ref{phox}. The following results were observed:

At halo mass proxies below $10^{13}$ $M_{\odot}$ (top three panels in Fig. \ref{stacking}), the IGM surface brightness profile is consistently overestimated by a factor that varies depending on the prior catalog. This effect is due to the lack of reliable calibration of the halo mass proxy at this scale and the large scatter in the correlation, as highlighted in Fig. \ref{fig2} and Table \ref{table1}. Results based on the R11 prior sample exhibit a larger discrepancy, particularly in the $10^{12.8}-10^{13.1}$ $M_{\odot}$ halo mass proxy bin. This is likely due to significant contamination from higher mass halos in the bin defined by the halo mass proxy, as shown in Table \ref{table1}.

At halo mass proxies above $10^{13}$ $M_{\odot}$ (bottom three panels in Fig. \ref{stacking}), the IGM surface brightness profile is in remarkable agreement with the input surface brightness profile across all three prior samples. We do not observe any significant discrepancies due to the halo mass proxy at this mass scale because the scatter in the proxy calibration is much smaller than at lower masses. Additionally, contamination by lower and higher mass halos decreases by a factor of 2 compared to the lower mass bins (see Table \ref{table1}), and the contamination tends to be more symmetrical to the bin center (see Fig. \ref{fig2}).

\subsection{The stacking results}

We test the efficiency of the stacking procedure in retrieving the input IGM surface brightness profile (see Fig. \ref{phox}) using the three optically selected group samples analyzed here. The stacking is performed at the coordinates of the optically selected group samples and within halo mass bins determined by each sample's halo mass proxy. The X-ray emissivity is based on the temperature proxy derived from the mass-temperature relation, as described above, assuming a $0.3Z_{\odot}$ metallicity. The effects of the assumed temperature and metallicity are included in the error budget by summing in quadrature the error estimated via bootstrapping and the effects estimated at different halo mass bins in previous sections. AGN contamination is removed by following the predictions from the simulation estimated in Table \ref{table2}, while XRB contamination is removed using the \cite{lehmer2016} scaling relation in the 0.5-2 keV band. The effect of using the optical halo mass proxy is not corrected for, but it will be discussed here.

Figure \ref{after_stacking} shows the comparison between the retrieved stacked IGM X-ray surface brightness profiles based on the three different optical prior catalogs and the input average PHOX IGM profile obtained in bins of input halo mass as in Fig. \ref{fig3}.

Despite the large error bars due to the aforementioned effects and the background subtraction, particularly at radii larger than $\sim0.7-0.9r_{200}$, the stacked profiles exhibit remarkably good agreement. The only evident effect is an overestimation of the R11-based stacked profile in the halo mass proxy bin at $10^{12.8-13.1}$ $M_{\odot}$, which was also observed in Fig. \ref{stacking} and is due to the large contamination of high mass halos at halo masses larger than the bin's upper limit.

The results demonstrate that the stacking procedure can accurately reproduce the average input profile derived from averaging the PHOX profiles of the underlying halo population at the same mass. This implies that:

\begin{itemize}
\item {\it{i)}} The optical selection effectively captures the average properties of the parent halo population.
\item {\it{ii)}} The stacking procedure performs proper background subtraction and accurately captures the shape of the input X-ray surface brightness profile up to $r_{200}$.
\item {\it{iii)}} The group center location does not statistically affect the profile shape, as the uncertainty is significantly smaller than the FWHM of the PSF, and any differences are mitigated by PSF convolution.
\item {\it{iv)}} The assumed temperature and metallicity do not significantly affect the derived profile but should be considered in the error budget.
\item {\it{v)}} Ignoring any temperature and metallicity spatial profiles does not dramatically affect the shape of the retrieved X-ray surface brightness profile.
\item {\it{vi)}} The main source of uncertainty derives from using the optical halo mass proxy, particularly when contamination is significant. Nevertheless, this seems to affect only a single bin and a single prior catalog.
\end{itemize}

Therefore, we conclude that utilizing the optically selected sample as a prior catalog for the stacking procedure defined in \cite{Popesso24} yields a reliable and robust estimate of the average X-ray surface brightness profile of galaxy groups when all selection and contamination effects are considered.

\begin{figure*}
\includegraphics[width=\textwidth]{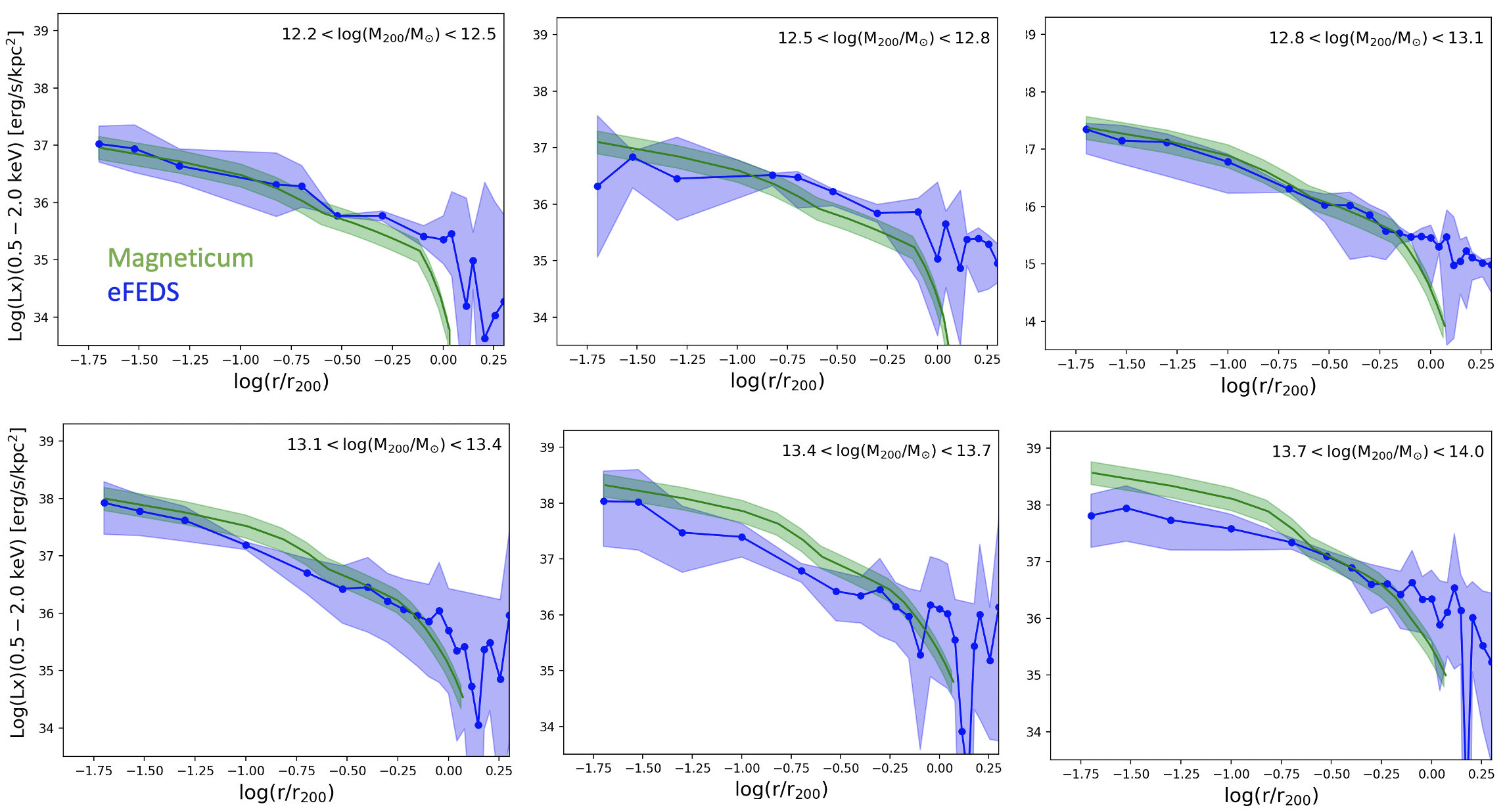}
\caption{The figure shows the comparison in different halo mass bins between the PHOX input IGM profiles convolved with the eROSITA PSF (green-shaded region) and the observed stacked profiles of \citet[][blue-shaded region]{Popesso2024c}. The latter profiles are obtained by following the same stacking procedure used here and with the real R11 GAMA optically selected group catalog. The shaded regions indicate the $1\sigma$ uncertainty.}
\label{mio}
\end{figure*}

\begin{figure*}
\includegraphics[width=\textwidth]{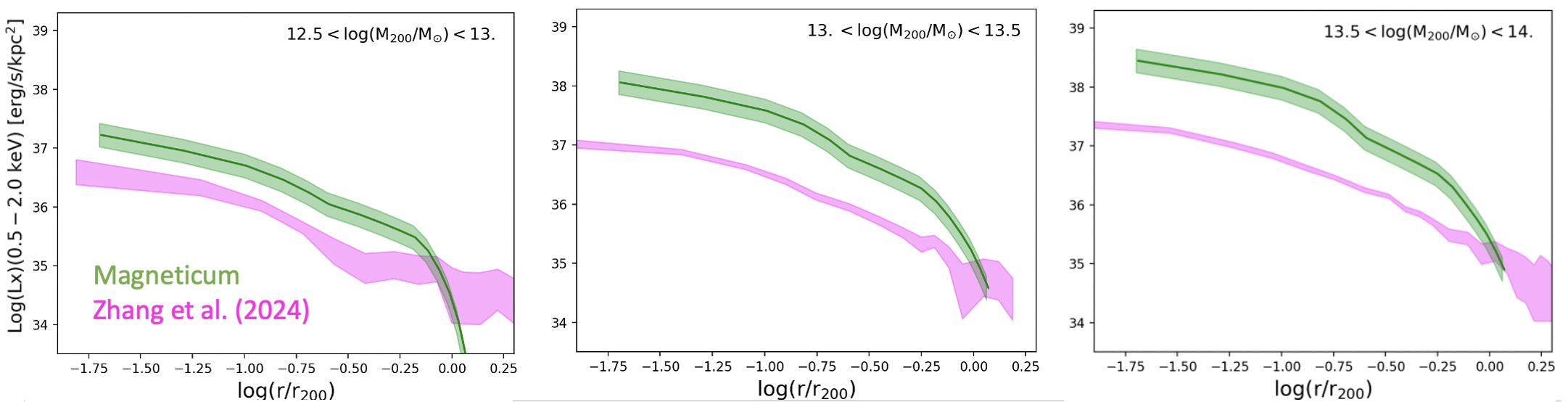}
\caption{The figure shows the comparison in different halo mass bins between the PHOX IGM input profiles convolved with the eROSITA PSF and the observed stacked profiles of \cite[][magenta-shaded region]{Zhang24a}. The latter profiles are obtained by following the stacking procedure outlined in \cite{Zhang24a} and with the real \cite{tinker2021} optically selected group catalog in the SDSS area. The shaded regions indicate the $1\sigma$ uncertainty.}
\label{zhang}
\end{figure*}

\subsection{The reconstruction of the gas mass profile}

The X-ray surface brightness profile carries information on how the hot gas is distributed in halos. The hot gas mass profile is usually reconstructed from the X-ray surface brightness profile, by assuming and electron density model. Here we use the \citet{Vikhlinin06} electron number density model, as usually done in the literature \citep[see also][]{LiuAng2022, Bulbul24}:

\begin{equation}
n_{\mathrm e}^2(r) = n_0^2 \cdot \left( \frac{r}{r_{\mathrm c}} \right)^{-\alpha} \cdot \left( 1 + \left( \frac{r}{r_{\mathrm c}} \right)^2 \right)^{-3\beta+\alpha/2} \cdot \left(  1 + \left( \frac{r}{r_{\mathrm s}} \right)^3 \right)^{-\epsilon/3},
\end{equation}

\noindent where $n_0$ is the normalization factor, $r_{\mathrm c}$ and $r_{\mathrm s}$ are the core and scale radii, $\beta$ controls the overall slope of the density profile, $\alpha$ controls the slope in the core and at intermediate radii, and $\epsilon$ controls the change of slope at large radii. This is used to estimate and integrate along the line of sight the X-ray emissivity,  $n_e(r)^2\Lambda(kT, Z)$ profile, where $\Lambda(kT, Z)$ is the cooling function depending on the gas temperature and metallicity. To mimic the observational technique, we estimate the cooling function by assuming a gas temperature from the $M-T_X$ relation of \cite{Lovisari2015} at the mean $M_{500}$ of the halo mass (proxy) bin. As in \citet{LiuAng2022}, we assumed a metallicity of the ICM of $0.3\, Z_{\odot}$, adopting the solar abundance table of \citet{asplund09}, that includes the He abundance. We neglect any temperature and metallicity profile. The projected number density model is convolved with the eROSITA PSF and fitted to the data to retrieve the best fit parameters. 

Figure \ref{ne_profile} presents the electron density profile predicted by the Magneticum simulation for a random subsample of halos across six halo mass bins. This is compared to the best-fit electron number density model from \citet{Vikhlinin06}, derived from fitting the stacked X-ray surface brightness profiles in Figure \ref{after_stacking}. We display the best-fit model for each mock, optically selected galaxy group sample. Despite substantial uncertainties in emissivity due to assumptions about mean gas temperature and metallicity, the profile obtained through the stacking method accurately reproduces the input average profiles.

\section{Comparison with observed IGM profiles}
Here we provide a comparison between the predicted Magneticum IGM profiles and the one recently obtained from the eROSITA survey and based on the stacking analysis. 

Fig. \ref{mio} presents the comparison between the average PHOX IGM profiles of Fig. \ref{phox} and the observed stacked profiles obtained in \citet{Popesso2024c} by following the same stacking technique tested here and based on the real R11 GAMA optically selected group sample. The observed stacked profiles are obtained in halo mass bins based on the total luminosity halo mass proxy tested here. The AGN and XRB contamination are corrected also in the real data as explained in Section 4.2. In all cases, the predicted PHOX profiles agree within 1$\sigma$ with the observed profiles.

Fig. \ref{zhang} presents a comparison of the PHOX IGM profiles (shown in Fig. \ref{phox}) with the observed stacked profiles from \cite{Zhang24a}, derived from the optically selected group catalog of \cite[][hereafter T21]{tinker2021}. As shown in Fig. \ref{zhang}, the observed profiles do not align with either the Magneticum predictions or the results from \citet{Popesso2024c}. The primary cause of this discrepancy is the use of different halo mass proxies in the T21 and R11 optical group prior catalogs. Despite both methods applying corrections for AGN contamination and satellite X-ray background (XRB), the underlying issue is a systematic underestimation of X-ray flux in the stacked profiles prior to these corrections, as highlighted in \cite{Zhang24a}. This discrepancy is likely attributable to differences in how the halo mass proxy is defined and estimated.

To investigate this further, we compare the T21 halo mass proxy with the Y05 estimates over the SDSS area, as both algorithms use the same dataset (see Appendix). Both rely on total optical luminosity to estimate halo mass via a calibration, but T21 adopts a more complex calibration that depends on galaxy properties (e.g., color), whereas Y05 uses a single power law. A direct comparison of the T21 and Y05 halo mass proxies reveals discrepancies in both the total luminosity estimates and the resulting halo masses (see Fig. \ref{tinker}). Given that \citet{Marini24b} extensively tested the Y05 algorithm for accurately identifying central and member galaxies and reliably estimating halo masses, we conclude that misidentifications in group membership, coupled with a less accurately tested calibration in T21, likely lead to an over-representation of low-mass, low-X-ray luminosity groups in the T21 catalog at fixed halo masses (see \ref{tinker_mass}). This, in turn, results in a reduced mean X-ray flux in the stacked profiles.

This finding emphasizes the importance of carefully considering the halo mass proxy and selection biases in the prior catalog, as these factors significantly influence the outcomes of X-ray stacking analyses (see Appendix for further details).

\section{Scaling relations}

\subsection{The $L_X$-Mass relation down to MW sized halos}

\begin{figure}
\includegraphics[width=0.5\textwidth]{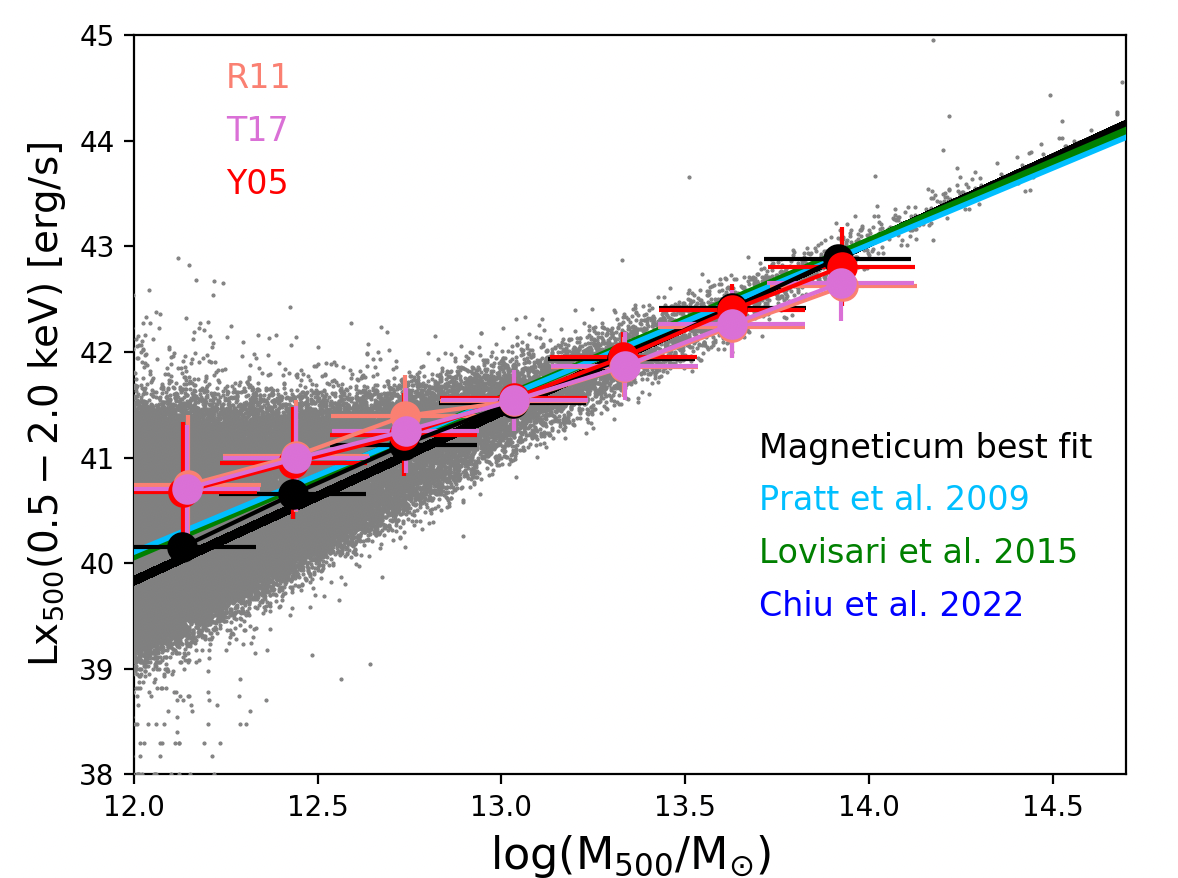}
\caption{$L_X-$mass relation in the $0.5-2.0$ keV band based on the stacking analysis applied to the Magneticum-eROSITA mock observations and based on the R11, T17 ans Y05 optically selected galaxy group catalog. The relations are color-coded as outlined in the figure: orange for R11, magenta for T17, and red for Y05 prior catalogs, respectively. The black symbols indicate the mean relation of the input Magneticum halos, while the black solid line indicates the best fit obtained by considering input X-ray luminosity and mass for halos with $M_{500} > 10^{13.5}$ $M_{\odot}$. The cyan solid line indicates the relation of \cite{pratt09}. The green solid line shows the relation of \cite{Lovisari2015} and the blue line the one of \cite{Chiu2022} in the $0.5-2.0$ keV band.}
\label{lx_mass}
\end{figure}

We present the predicted $L_X$-Mass relation based on the stacking analysis of the Magneticum-eROSITA mock observation, using the three optically selected prior catalogs considered in this study. \cite{Marini2024a} have already shown that the $L_X$-Mass relation predicted by Magneticum is consistent with the available observational relations, calibrated down to halo masses of $M_{500} \sim 10^{13.5}$ $M_{\odot}$ \citep[see e.g.][]{Pratt2010, Lovisari2015, Chiu2022}. Here, we demonstrate in Fig \ref{lx_mass} that this relation can be accurately retrieved by our stacking analysis. The stacked relations were derived by integrating the stacked profiles from Fig. \ref{after_stacking} within $r_{500}$. Additionally, we include a data point for clusters with $M_{500} > 10^{14}$ $M_{\odot}$. Nearly all optically selected clusters have been detected as eSASS extended sources.

All stacked points, irrespective of the group selection algorithm, result in flatter $L_X$-Mass relations compared to the input relation, with the greatest discrepancies observed in the lowest halo mass bins, as shown in Fig. \ref{after_stacking}. According to our analysis of potential sources of systematic errors and uncertainties discussed in the previous sections, this discrepancy is primarily due to the halo mass proxy used for binning the prior samples. Specifically, we employed the same proxy across all three catalogs, the group total optical luminosity, which is poorly calibrated for halo masses below $10^{13}$ $M_{\odot}$. Any systematic errors in this calibration (see Table \ref{table1}) propagate as systematic errors in the $L_X$-Mass slope. Therefore, caution should be exercised when comparing the $L_X$-Mass slope obtained from stacked scaling relations with other studies in the literature that utilize different halo mass proxies and predictions.

\subsection{The $f_{gas}-M_{500}$ relation down to MW sized halos}

The $f_{gas}-M_{500}$ relation is a key to understand how much hot gas is locked in massive halos within a given radius. Here we focus on the relation estimated within $r_{500}$. The mass of the hot gas within each halo mass bin is estimated by integrating the best-fit electron number density model from \citet{Vikhlinin06} obtained in Fig. \ref{ne_profile}. The gas mass within $r_{500}$ is normalized to the $M_{500}$ estimate based on mass proxy of the corresponding optically selected sample used as prior for the stacking. The predicted relation from Magneticum is, instead, calculated by averaging the predicted hot gas mass fraction within a halo mass bin. This is estimated by considering all gas particle with a tempertaure higher than $10^5$ K. 

\begin{figure}
\includegraphics[width=0.5\textwidth]{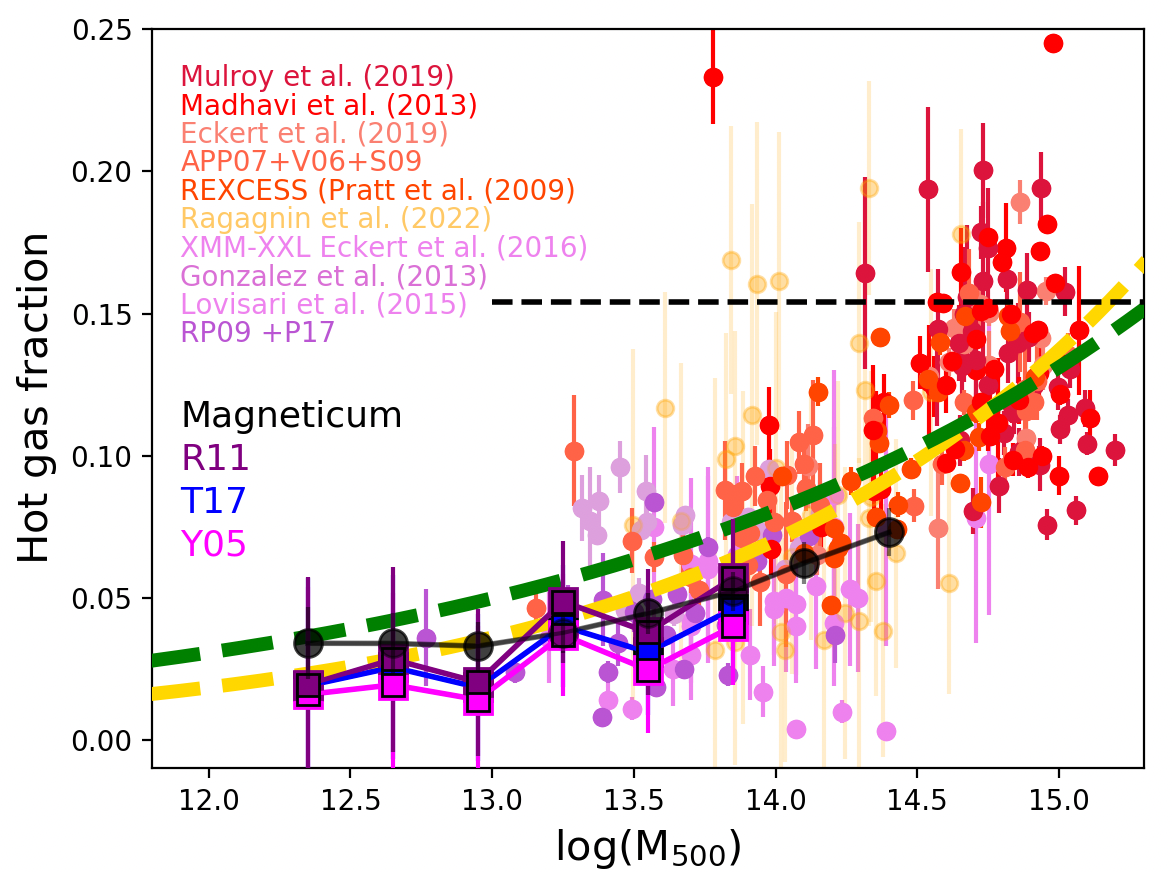}
\caption{$f_\mathrm{gas}$-$M_{500}$ relation. The filled circles represent a compilation of literature data, color-coded as a function of the sample reference as shown in the picture. The black points connected by a solid line represent the Magneticum prediction. The squares indicate the $f_\mathrm{gas}$ derived from the stacked X-ray surface brightness profiles. This is obtained by integrating the best fit electron density model of Fig.\ref{ne_profile}. The dashed yellow and green curves represent the best fit of \citet{pratt09} and \citet{Eckert16}, respectively. The dashed line indicates the value of the cosmic baryon fraction according to the WMAP cosmology implemented in Magneticum.}
\label{hot_gas}
\end{figure}

Fig.\ref{hot_gas} shows the comparison between the Magneticum prediction and the relation based in the stacked profiles. The error bars for the Magneticum prediction represent the dispersion around the mean, while the error bars of the gas fraction based on the stacks are obtained by summing in quadrature the error of the best fit due to the observed error in the stacked surface brightness profiles and the uncertainty due to the assumption of the mean gas temperature and metallicity in the emissivity (see Section \ref{emissivity_1}). The input predictions and those retrive by applying the observational technique to the mock dataset are perfectly in agreement within the error bars. This ensures that the stacking technique is capable of retrieving the input average properties of the underlying halo population.

In addition, in Fig. \ref{hot_gas} we also show a comprehensive compilation of literature data providing the same estimate for $z < 0.2$ systems. These are the cluster sample of \citet{Mulroy19}, \citet{Madhavi13}, \citet{eckert19}, the REXCESS sample of \citet{pratt09}, the poor cluster and group sample of \citet{Arnaud07}, \citet{Vikhlinin06}, and \citet{Sun2009} used in \citet{pratt09} (APP05$+$V09$+$S09 in Fig.\ref{hot_gas}), the XMM-XXL sample of \citet{Eckert16}, the group samples of \citet{Gonzalez13} and \citet{Lovisari2015}, the poor cluster sample of \citet{Ragagnin22}, and the group sample of \citet{Rasmussen09} and \citet{Pearson17} (RP09$+$P17 in Fig.\ref{hot_gas}). We transformed all the values into our adopted cosmology. Given the very different selection functions of the collected data, it is not surprising that the scatter is large. Nevertheless, we find consistency in the distribution of the observed systems and the mean prediction of Magneticum in the group and poor cluster regimes.

\section{Discussion and conclusions}
The creation of eROSITA mock observations at the eRASS:4 depth and optical galaxy catalogs based on the same Magneticum lightcones in the local Universe \citep[see][]{Marini2024a} allows us to generate synthetic analogs that mirror the available observational datasets. These datasets are created by combining eROSITA observations with optically selected catalogs such as those by \cite{Robotham2011} in the GAMA area, and \cite{Tempel2017} and \cite{Yang2005} in the SDSS region. We created synthetic analogs of optically selected group catalogs using the selection algorithms from these three reference papers.

We present a thorough analysis of the sources of systematic errors and uncertainties involved in the stacking analysis of optically selected groups in the eROSITA data. In particular, we test the impact of these systematics and uncertainties on the results of the stacking technique by \cite{popesso2019}, applied to the three synthetic optically selected group catalogs considered here. Following the findings of \cite{Marini2024a}, we use the total optical group luminosity as a halo mass proxy, which exhibits the least scatter when compared with the input halo mass. We discuss the following:

\begin{enumerate}
\item [-] The completeness and contamination of the prior optical group catalogs and the effect of any selection biases.
\item [-] The effect of possible optical miscentering when stacking at the coordinates of the optically selected groups.
\item [-] The impact on X-ray emissivity estimation when assuming a temperature based on the $T_X$-mass scaling relation through the halo mass proxy, and a fixed metallicity ($0.3Z_{\odot}$).
\item [-] The contamination from AGN and XRB emission.
\item [-] The effect of the scatter between the halo mass proxy and the input halo mass on the contamination within each halo mass bin.
\end{enumerate}

To test these effects, we construct predicted contributions to the X-ray emission in different halo mass bins and projected X-ray surface brightness profiles. We isolate the IGM contribution in each halo mass bin and evaluate how effectively the stacking procedure retrieves the input IGM surface brightness profile. This analysis includes halos with masses below $10^{13}$ $M_{\odot}$. However, we note that the mean virial temperature of the hot gas in such halos is below the energy range of the eROSITA soft band. Therefore, the provided predictions of the IGM profiles in these halo mass bins only reflect the gas particles in the high-temperature tail of the gas temperature distribution. Consequently, these predictions are not representative of the entire CGM or IGM emission in low-mass systems.

The analysis leads to the following results:

\begin{enumerate}
\item [-] The completeness and contamination of the prior optical group catalogs are such that no significant selection effects are observed.
\item [-] The differences in coordinates ($\Delta{RA}$ and $\Delta{Dec}$) between the input halo center and the optical center are much smaller than the eROSITA PSF at the considered redshifts. Therefore, any mis-centering effect is mitigated when the Magneticum projected PHOX X-ray surface brightness profiles are convolved with the eROSITA PSF. We do not observe significant mis-centering relative to the X-ray center in the eSASS extended source detections matched to the input and optically selected group catalogs.
\item [-] The effect on X-ray emissivity depends on the systematics of the halo mass proxy. If this proxy is used to infer the mean temperature of the gas from the $T_X$-mass relation, the temperature can be over- or underestimated. For the R11 catalog, the total luminosity proxy results in a scatter of 40\% in the emissivity estimate at fixed metallicity. For the T17 and Y05 catalogs, we observe an overestimation of approximately 20\% below 0.6 keV and an underestimation of 10\% above this threshold. The assumption of a fixed metallicity of $0.3Z_{\odot}$ aligns relatively well with the Magneticum simulation, as 85\% of the halos in the two orders of magnitude in halo mass considered here exhibit a mean metallicity in the range $0.1 - 0.45Z_{\odot}$, leading to a variation in emissivity of 37\%. Since these effects cannot be corrected when estimating the electron number density profile and gas mass, they are included in the error budget and summed in quadrature with the error due to background subtraction.
\item [-] According to the Magneticum predictions, AGN contamination dominates the X-ray surface brightness profile emission in all halos with masses below $10^{13}$ $M_{\odot}$, which contain the majority of low X-ray luminosity AGN. Consistent with \cite{Popesso24}, AGN contamination is limited to the activity of the central galaxy and is well modeled with a PSF rescaled to the percentage of the X-ray emission corresponding to the AGN contribution. Similarly, XRB contamination is mainly due to the central galaxy's activity and is significant on average only for halos with masses below $10^{13}$ $M_{\odot}$, while it can be neglected for more massive halos. This contamination can also be modeled as a PSF and subtracted from the total X-ray emission.
\item [-] The most significant systematic effect in the stacking analysis is the scatter between the halo mass proxy and the input halo mass. Each halo mass bin sample might be differently contaminated by high and low mass systems depending on the level of agreement between the halo mass proxy and the input mass. This systematic leads to an overall overestimation of the X-ray surface brightness profiles, particularly for halos with masses below $10^{13}$ $M_{\odot}$.
\item [-] When performing the stacking analysis in the mock observations, all these effects tend to compensate for each other to some extent. We confirm an overestimation of the X-ray surface brightness profile for low mass groups below $10^{13}$ $M_{\odot}$ mainly due to contamination in the bins of the halo mass proxy. For halos with masses above $10^{13}$ $M_{\odot}$, we find good agreement between the stacked and input projected PHOX profiles.
\item [-] Any systematic error in the calibration of the halo mass proxy is propagated when estimating the scaling relations based on the stacks. This is more visible in the $L_X-$mass relation. We show the relations based on the stacking of the different optically selected group priors and find that, in all cases, the slope of the relations is flatter than the input relation. Such systematic errors must be considered when comparing the results of any stacking technique with other works in the literature based on different prior catalogs, detections, or predictions.
\end{enumerate}

The comparison of the Magneticum input IGM PHOX profiles with the observed profiles of \citet{Popesso2024c} shows that the predictions align well with the observations. However, when these results are compared with the X-ray surface brightness profile from \cite{Zhang24a}, a discrepancy is noted across all considered halo masses, with the corresponding flux level being lower than that of Magneticum and the observed $L_X$-mass relation \citep[e.g.,][]{pratt09,Lovisari2015,Chiu2022}. This discrepancy highlights the impact of systematics in the stacking analysis, which can lead to differences when comparing results based on different prior samples and techniques.

We conclude that overall, the stacking technique is reliable when all systematics are properly considered and corrected where possible. If a correction is not feasible, the uncertainty should be included in the error budget. Once the selection effects of the optical selection algorithms are properly checked, the main source of systematics and uncertainty remains the halo mass proxy chosen for the binning in halo mass. Using the best proxy with the smallest scatter compared to the input halo mass, namely the total optical luminosity, still leads to some contamination in the lowest halo mass bins, which can affect the study of the slope of the $L_X$-mass relation.

Overall, Magneticum predictions are consistent with observations of scaling relations and can accurately reproduce the observed X-ray surface brightness profiles over the two decades of halo masses in the group regime below $10^{14}$ $M_{\odot}$.

\begin{acknowledgements}
PP acknowledges financial support from the European Research Council (ERC) under the European Union’s Horizon Europe research and innovation programme ERC CoG CLEVeR (Grant agreement No. 101045437).
AB acknowledges the financial contribution from the INAF mini-grant 1.05.12.04.01 {\it "The dynamics of clusters of galaxies from the projected phase-space distribution of cluster galaxies"}. KD acknowledges support by the COMPLEX project from the European Research Council (ERC) under the European Union’s Horizon 2020 research and innovation program grant agreement ERC-2019-AdG 882679. The calculations for the {\it Magneticum} simulations were carried out at the Leibniz Supercomputer Center (LRZ) under the project pr83li. 
GP acknowledges financial support from the European Research Council (ERC) under the European Union’s Horizon 2020 research and innovation program Hot- Milk (grant agreement No 865637), support from Bando per il Finanziamento della Ricerca Fondamentale 2022 dell’Istituto Nazionale di Astrofisica (INAF): GO Large program and from the Framework per l’Attrazione e il Rafforzamento delle Eccellenze (FARE) per la ricerca in Italia (R20L5S39T9). SVZ and VB acknowledge support by the \emph{Deut\-sche For\-schungs\-ge\-mein\-schaft, DFG\/} project nr. 415510302

This work is based on data from eROSITA, the soft X-ray instrument aboard SRG, a joint Russian-German science mission supported by the Russian Space Agency (Roskosmos), in the interests of the Russian Academy of Sciences represented by its Space Research Institute (IKI), and the Deutsches Zentrum für Luft- und Raumfahrt (DLR). The SRG spacecraft was built by Lavochkin Association (NPOL) and its subcontractors and is operated by NPOL with support from the Max Planck Institute for Extraterrestrial Physics (MPE).

The development and construction of the eROSITA X-ray instrument was led by MPE, with contributions from the Dr. Karl Remeis Observatory Bamberg \& ECAP (FAU Erlangen-Nuernberg), the University of Hamburg Observatory, the Leibniz Institute for Astrophysics Potsdam (AIP), and the Institute for Astronomy and Astrophysics of the University of Tübingen, with the support of DLR and the Max Planck Society. The Argelander Institute for Astronomy of the University of Bonn and the Ludwig Maximilians Universität Munich also participated in the science preparation for eROSITA.

The eROSITA data shown here were processed using the eSASS software system developed by the German eROSITA consortium.

\end{acknowledgements}

%
%

\bibliographystyle{aa} 
\bibliography{biblio.bib} 

\begin{thebibliography}{78}
\expandafter\ifx\csname natexlab\endcsname\relax\def\natexlab#1{#1}\fi

\bibitem[{{Aird} \& {Coil}(2021)}]{Aird2021}
{Aird}, J. \& {Coil}, A.~L. 2021, \mnras, 502, 5962

\bibitem[{{Alpaslan} \& {Tinker}(2020)}]{Alpaslan2020}
{Alpaslan}, M. \& {Tinker}, J.~L. 2020, \mnras, 496, 5463

\bibitem[{{Anderson} {et~al.}(2015){Anderson}, {Gaspari}, {White}, {Wang}, \& {Dai}}]{Anderson2015}
{Anderson}, M.~E., {Gaspari}, M., {White}, S. D.~M., {Wang}, W., \& {Dai}, X. 2015, \mnras, 449, 3806

\bibitem[{{Arnaud} {et~al.}(2007){Arnaud}, {Pointecouteau}, \& {Pratt}}]{Arnaud07}
{Arnaud}, M., {Pointecouteau}, E., \& {Pratt}, G.~W. 2007, \aap, 474, L37

\bibitem[{{Asplund} {et~al.}(2009){Asplund}, {Grevesse}, {Sauval}, \& {Scott}}]{asplund09}
{Asplund}, M., {Grevesse}, N., {Sauval}, A.~J., \& {Scott}, P. 2009, \araa, 47, 481

\bibitem[{Beck {et~al.}(2016)Beck, Murante, Arth, Remus, Teklu, Donnert, Planelles, Beck, Förster, Imgrund, Dolag, \& Borgani}]{beck_improved_2016}
Beck, A.~M., Murante, G., Arth, A., {et~al.} 2016, MNRAS, 455, 2110

\bibitem[{{Behroozi} {et~al.}(2019){Behroozi}, {Wechsler}, {Hearin}, \& {Conroy}}]{Berhoozi2019}
{Behroozi}, P., {Wechsler}, R.~H., {Hearin}, A.~P., \& {Conroy}, C. 2019, \mnras, 488, 3143

\bibitem[{Biffi {et~al.}(2013)Biffi, Dolag, \& Böhringer}]{biffi_investigating_2013}
Biffi, V., Dolag, K., \& Böhringer, H. 2013, MNRAS, 428, 1395

\bibitem[{Biffi {et~al.}(2012)Biffi, Dolag, Böhringer, \& Lemson}]{biffi_observing_2012}
Biffi, V., Dolag, K., Böhringer, H., \& Lemson, G. 2012, MNRAS, 420, 3545

\bibitem[{{Biffi} {et~al.}(2018){Biffi}, {Dolag}, \& {Merloni}}]{Biffi18}
{Biffi}, V., {Dolag}, K., \& {Merloni}, A. 2018, \mnras, 481, 2213

\bibitem[{{Bulbul} {et~al.}(2024){Bulbul}, {Liu}, {Kluge}, {Zhang}, {Sanders}, {Bahar}, {Ghirardini}, {Artis}, {Seppi}, {Garrel}, {Ramos-Ceja}, {Comparat}, {Balzer}, {B{\"o}ckmann}, {Br{\"u}ggen}, {Clerc}, {Dennerl}, {Dolag}, {Freyberg}, {Grandis}, {Gruen}, {Kleinebreil}, {Krippendorf}, {Lamer}, {Merloni}, {Migkas}, {Nandra}, {Pacaud}, {Predehl}, {Reiprich}, {Schrabback}, {Veronica}, {Weller}, \& {Zelmer}}]{Bulbul24}
{Bulbul}, E., {Liu}, A., {Kluge}, M., {et~al.} 2024, arXiv e-prints, arXiv:2402.08452

\bibitem[{{Chiu} {et~al.}(2022){Chiu}, {Ghirardini}, {Liu}, {Grandis}, {Bulbul}, {Bahar}, {Comparat}, {Bocquet}, {Clerc}, {Klein}, {Liu}, {Li}, {Miyatake}, {Mohr}, {More}, {Oguri}, {Okabe}, {Pacaud}, {Ramos-Ceja}, {Reiprich}, {Schrabback}, \& {Umetsu}}]{Chiu2022}
{Chiu}, I.~N., {Ghirardini}, V., {Liu}, A., {et~al.} 2022, \aap, 661, A11

\bibitem[{{Clerc} {et~al.}(2018){Clerc}, {Ramos-Ceja}, {Ridl}, {Lamer}, {Brunner}, {Hofmann}, {Comparat}, {Pacaud}, {K{\"a}fer}, {Reiprich}, {Merloni}, {Schmid}, {Brand}, {Wilms}, {Friedrich}, {Finoguenov}, {Dauser}, \& {Kreykenbohm}}]{Clerc2018}
{Clerc}, N., {Ramos-Ceja}, M.~E., {Ridl}, J., {et~al.} 2018, \aap, 617, A92

\bibitem[{{Comparat} {et~al.}(2023){Comparat}, {Luo}, {Merloni}, {More}, {Salvato}, {Krumpe}, {Miyaji}, {Brandt}, {Georgakakis}, {Akiyama}, {Buchner}, {Dwelly}, {Kawaguchi}, {Liu}, {Nagao}, {Nandra}, {Silverman}, {Toba}, {Anderson}, \& {Kollmeier}}]{comparat2023}
{Comparat}, J., {Luo}, W., {Merloni}, A., {et~al.} 2023, \aap, 673, A122

\bibitem[{{Comparat} {et~al.}(2022){Comparat}, {Truong}, {Merloni}, {Pillepich}, {Ponti}, {Driver}, {Bellstedt}, {Liske}, {Aird}, {Br{\"u}ggen}, {Bulbul}, {Davies}, {Gonz{\'a}lez Villalba}, {Georgakakis}, {Haberl}, {Liu}, {Maitra}, {Nandra}, {Popesso}, {Predehl}, {Robotham}, {Salvato}, {Thorne}, \& {Zhang}}]{Comparat2022}
{Comparat}, J., {Truong}, N., {Merloni}, A., {et~al.} 2022, arXiv e-prints, arXiv:2201.05169

\bibitem[{Dauser {et~al.}(2019)Dauser, Falkner, Lorenz, Kirsch, Peille, Cucchetti, Schmid, Brand, Oertel, Smith, \& Wilms}]{dauser_sixte_2019}
Dauser, T., Falkner, S., Lorenz, M., {et~al.} 2019, A\& A, 630, A66

\bibitem[{Di~Matteo {et~al.}(2005)Di~Matteo, Springel, \& Hernquist}]{di_matteo_energy_2005}
Di~Matteo, T., Springel, V., \& Hernquist, L. 2005, Nature, 433, 604

\bibitem[{Dolag {et~al.}(2009)Dolag, Borgani, Murante, \& Springel}]{dolag_substructures_2009}
Dolag, K., Borgani, S., Murante, G., \& Springel, V. 2009, MNRAS, 399, 497

\bibitem[{{Dolag} {et~al.}(2016){Dolag}, {Komatsu}, \& {Sunyaev}}]{Dolag16}
{Dolag}, K., {Komatsu}, E., \& {Sunyaev}, R. 2016, \mnras, 463, 1797

\bibitem[{Dolag {et~al.}(2005)Dolag, Vazza, Brunetti, \& Tormen}]{dolag_turbulent_2005}
Dolag, K., Vazza, F., Brunetti, G., \& Tormen, G. 2005, MNRAS, 364, 753

\bibitem[{{Eckert} {et~al.}(2016){Eckert}, {Ettori}, {Coupon}, {Gastaldello}, {Pierre}, {Melin}, {Le Brun}, {McCarthy}, {Adami}, {Chiappetti}, {Faccioli}, {Giles}, {Lavoie}, {Lef{\`e}vre}, {Lieu}, {Mantz}, {Maughan}, {McGee}, {Pacaud}, {Paltani}, {Sadibekova}, {Smith}, \& {Ziparo}}]{Eckert16}
{Eckert}, D., {Ettori}, S., {Coupon}, J., {et~al.} 2016, \aap, 592, A12

\bibitem[{{Eckert} {et~al.}(2021){Eckert}, {Gaspari}, {Gastaldello}, {Le Brun}, \& {O'Sullivan}}]{2021Univ....7..142E}
{Eckert}, D., {Gaspari}, M., {Gastaldello}, F., {Le Brun}, A. M.~C., \& {O'Sullivan}, E. 2021, Universe, 7, 142

\bibitem[{{Eckert} {et~al.}(2019){Eckert}, {Ghirardini}, {Ettori}, {Rasia}, {Biffi}, {Pointecouteau}, {Rossetti}, {Molendi}, {Vazza}, {Gastaldello}, {Gaspari}, {De Grandi}, {Ghizzardi}, {Bourdin}, {Tchernin}, \& {Roncarelli}}]{eckert19}
{Eckert}, D., {Ghirardini}, V., {Ettori}, S., {et~al.} 2019, \aap, 621, A40

\bibitem[{{Erfanianfar} {et~al.}(2019){Erfanianfar}, {Finoguenov}, {Furnell}, {Popesso}, {Biviano}, {Wuyts}, {Collins}, {Mirkazemi}, {Comparat}, {Khosroshahi}, {Nandra}, {Capasso}, {Rykoff}, {Wilman}, {Merloni}, {Clerc}, {Salvato}, {Chitham}, {Kelvin}, {Gozaliasl}, {Weijmans}, {Brownstein}, {Egami}, {Pereira}, {Schneider}, {Kirkpatrick}, {Damsted}, \& {Kukkola}}]{ghazaleh1}
{Erfanianfar}, G., {Finoguenov}, A., {Furnell}, K., {et~al.} 2019, \aap, 631, A175

\bibitem[{{Erfanianfar} {et~al.}(2014){Erfanianfar}, {Popesso}, {Finoguenov}, {Wuyts}, {Wilman}, {Biviano}, {Ziparo}, {Salvato}, {Nandra}, {Lutz}, {Elbaz}, {Dickinson}, {Tanaka}, {Mirkazemi}, {Balogh}, {Altieri}, {Aussel}, {Bauer}, {Berta}, {Bielby}, {Brandt}, {Cappelluti}, {Cimatti}, {Cooper}, {Fadda}, {Ilbert}, {Le Floch}, {Magnelli}, {Mulchaey}, {Nordon}, {Newman}, {Poglitsch}, \& {Pozzi}}]{ghazaleh2}
{Erfanianfar}, G., {Popesso}, P., {Finoguenov}, A., {et~al.} 2014, \mnras, 445, 2725

\bibitem[{Fabjan {et~al.}(2010)Fabjan, Borgani, Tornatore, Saro, Murante, \& Dolag}]{fabjan_simulating_2010}
Fabjan, D., Borgani, S., Tornatore, L., {et~al.} 2010, MNRAS, 401, 1670

\bibitem[{{Georgakakis} {et~al.}(2019){Georgakakis}, {Comparat}, {Merloni}, {Ciesla}, {Aird}, \& {Finoguenov}}]{Georgakakis2019}
{Georgakakis}, A., {Comparat}, J., {Merloni}, A., {et~al.} 2019, \mnras, 487, 275

\bibitem[{{Gonzalez} {et~al.}(2013){Gonzalez}, {Sivanandam}, {Zabludoff}, \& {Zaritsky}}]{Gonzalez13}
{Gonzalez}, A.~H., {Sivanandam}, S., {Zabludoff}, A.~I., \& {Zaritsky}, D. 2013, \apj, 778, 14

\bibitem[{Haardt \& Madau(2001)}]{haardt_modelling_2001}
Haardt, F. \& Madau, P. 2001 (eprint: arXiv:astro-ph/0106018)

\bibitem[{Hirschmann {et~al.}(2014)Hirschmann, Dolag, Saro, Bachmann, Borgani, \& Burkert}]{hirschmann_cosmological_2014}
Hirschmann, M., Dolag, K., Saro, A., {et~al.} 2014, MNRAS, 442, 2304

\bibitem[{{Krumpe} {et~al.}(2018){Krumpe}, {Miyaji}, {Coil}, \& {Aceves}}]{krumpe2018}
{Krumpe}, M., {Miyaji}, T., {Coil}, A.~L., \& {Aceves}, H. 2018, \mnras, 474, 1773

\bibitem[{{Krumpe} {et~al.}(2023){Krumpe}, {Miyaji}, {Georgakakis}, {Schulze}, {Coil}, {Dwelly}, {Coffey}, {Comparat}, {Aceves}, {Salvato}, {Merloni}, {Maraston}, {Nandra}, {Brownstein}, {Schneider}, {SDSS-Iv Team}, \& {Spiders Team}}]{krumpe2023}
{Krumpe}, M., {Miyaji}, T., {Georgakakis}, A., {et~al.} 2023, \apj, 952, 109

\bibitem[{{Le Brun} {et~al.}(2014){Le Brun}, {McCarthy}, {Schaye}, \& {Ponman}}]{LeBrun2014}
{Le Brun}, A. M.~C., {McCarthy}, I.~G., {Schaye}, J., \& {Ponman}, T.~J. 2014, \mnras, 441, 1270

\bibitem[{{Lehmer} {et~al.}(2016){Lehmer}, {Basu-Zych}, {Mineo}, {Brandt}, {Eufrasio}, {Fragos}, {Hornschemeier}, {Luo}, {Xue}, {Bauer}, {Gilfanov}, {Ranalli}, {Schneider}, {Shemmer}, {Tozzi}, {Trump}, {Vignali}, {Wang}, {Yukita}, \& {Zezas}}]{lehmer2016}
{Lehmer}, B.~D., {Basu-Zych}, A.~R., {Mineo}, S., {et~al.} 2016, \apj, 825, 7

\bibitem[{{Liu} {et~al.}(2022{\natexlab{a}}){Liu}, {Bulbul}, {Ghirardini}, {Liu}, {Klein}, {Clerc}, {{\"O}zsoy}, {Ramos-Ceja}, {Pacaud}, {Comparat}, {Okabe}, {Bahar}, {Biffi}, {Brunner}, {Br{\"u}ggen}, {Buchner}, {Ider Chitham}, {Chiu}, {Dolag}, {Gatuzz}, {Gonzalez}, {Hoang}, {Lamer}, {Merloni}, {Nandra}, {Oguri}, {Ota}, {Predehl}, {Reiprich}, {Salvato}, {Schrabback}, {Sanders}, {Seppi}, \& {Thibaud}}]{LiuAng2022}
{Liu}, A., {Bulbul}, E., {Ghirardini}, V., {et~al.} 2022{\natexlab{a}}, \aap, 661, A2

\bibitem[{{Liu} {et~al.}(2022{\natexlab{b}}){Liu}, {Buchner}, {Nandra}, {Merloni}, {Dwelly}, {Sanders}, {Salvato}, {Arcodia}, {Brusa}, {Wolf}, {Georgakakis}, {Boller}, {Krumpe}, {Lamer}, {Waddell}, {Urrutia}, {Schwope}, {Robrade}, {Wilms}, {Dauser}, {Comparat}, {Toba}, {Ichikawa}, {Iwasawa}, {Shen}, \& {Medel}}]{LiuTeng2022}
{Liu}, T., {Buchner}, J., {Nandra}, K., {et~al.} 2022{\natexlab{b}}, \aap, 661, A5

\bibitem[{{Lovisari} \& {Ettori}(2021)}]{2021Univ....7..254L}
{Lovisari}, L. \& {Ettori}, S. 2021, Universe, 7, 254

\bibitem[{{Lovisari} \& {Reiprich}(2019)}]{lovisari19}
{Lovisari}, L. \& {Reiprich}, T.~H. 2019, \mnras, 483, 540

\bibitem[{{Lovisari} {et~al.}(2015){Lovisari}, {Reiprich}, \& {Schellenberger}}]{Lovisari2015}
{Lovisari}, L., {Reiprich}, T.~H., \& {Schellenberger}, G. 2015, \aap, 573, A118

\bibitem[{{Mahdavi} {et~al.}(2013){Mahdavi}, {Hoekstra}, {Babul}, {Bildfell}, {Jeltema}, \& {Henry}}]{Madhavi13}
{Mahdavi}, A., {Hoekstra}, H., {Babul}, A., {et~al.} 2013, \apj, 767, 116

\bibitem[{{Mandelbaum} {et~al.}(2016){Mandelbaum}, {Wang}, {Zu}, {White}, {Henriques}, \& {More}}]{Mandelbaum2016}
{Mandelbaum}, R., {Wang}, W., {Zu}, Y., {et~al.} 2016, \mnras, 457, 3200

\bibitem[{Marchesi {et~al.}(2020)Marchesi, Gilli, Lanzuisi, Dauser, Ettori, Vito, Cappelluti, Comastri, Mushotzky, Ptak, \& Norman}]{marchesi_mock_2020}
Marchesi, S., Gilli, R., Lanzuisi, G., {et~al.} 2020, A\&A, 642, A184

\bibitem[{{Marini} {et~al.}(2024{\natexlab{a}}){Marini}, {Popesso}, {Lamer}, {Dolag}, {Biffi}, {Vladutescu-Zopp}, {Dev}, {Toptun}, {Bulbul}, {Comparat}, {Malavasi}, {Merloni}, {Mroczkowski}, {Ponti}, {Seppi}, {Shreeram}, \& {Zhang}}]{Marini24b}
{Marini}, I., {Popesso}, P., {Lamer}, G., {et~al.} 2024{\natexlab{a}}, A\&A submitted

\bibitem[{{Marini} {et~al.}(2024{\natexlab{b}}){Marini}, {Popesso}, {Lamer}, {Dolag}, {Biffi}, {Vladutescu-Zopp}, {Dev}, {Toptun}, {Bulbul}, {Comparat}, {Malavasi}, {Merloni}, {Mroczkowski}, {Ponti}, {Seppi}, {Shreeram}, \& {Zhang}}]{Marini2024a}
{Marini}, I., {Popesso}, P., {Lamer}, G., {et~al.} 2024{\natexlab{b}}, \aap, 689, A7

\bibitem[{{McCarthy} {et~al.}(2017){McCarthy}, {Schaye}, {Bird}, \& {Le Brun}}]{McCarthy2017}
{McCarthy}, I.~G., {Schaye}, J., {Bird}, S., \& {Le Brun}, A. M.~C. 2017, \mnras, 465, 2936

\bibitem[{Merloni {et~al.}(2024)Merloni, Lamer, Liu, Ramos-Ceja, Brunner, Bulbul, Dennerl, Doroshenko, Freyberg, Friedrich, Gatuzz, Georgakakis, Haberl, Igo, Kreykenbohm, Liu, Maitra, Malyali, Mayer, Nandra, Predehl, Robrade, Salvato, Sanders, Stewart, Tubín-Arenas, Weber, Wilms, Arcodia, Artis, Aschersleben, Avakyan, Aydar, Bahar, Balzer, Becker, Berger, Boller, Bornemann, Brüggen, Brusa, Buchner, Burwitz, Camilloni, Clerc, Comparat, Coutinho, Czesla, Dannhauer, Dauner, Dauser, Dietl, Dolag, Dwelly, Egg, Ehl, Freund, Friedrich, Gaida, Garrel, Ghirardini, Gokus, Grünwald, Grandis, Grotova, Gruen, Gueguen, Hämmerich, Hamaus, Hasinger, Haubner, Homan, Ider~Chitham, Joseph, Joyce, König, Kaltenbrunner, Khokhriakova, Kink, Kirsch, Kluge, Knies, Krippendorf, Krumpe, Kurpas, Li, Liu, Locatelli, Lorenz, Müller, Magaudda, Mannes, McCall, Meidinger, Michailidis, Migkas, Muñoz-Giraldo, Musiimenta, Nguyen-Dang, Ni, Olechowska, Ota, Pacaud, Pasini, Perinati, Pires, Pommranz, Ponti, Poppenhaeger, Pühlhofer, Rau,
  Reh, Reiprich, Roster, Saeedi, Santangelo, Sasaki, Schmitt, Schneider, Schrabback, Schuster, Schwope, Seppi, Serim, Shreeram, Sokolova-Lapa, Starck, Stelzer, Stierhof, Suleimanov, Tenzer, Traulsen, Trümper, Tsuge, Urrutia, Veronica, Waddell, Willer, Wolf, Yeung, Zainab, Zangrandi, Zhang, Zhang, \& Zheng}]{merloni_srgerosita_2024}
Merloni, A., Lamer, G., Liu, T., {et~al.} 2024, A\&A, 682, A34

\bibitem[{{Mernier} {et~al.}(2017){Mernier}, {de Plaa}, {Kaastra}, {Zhang}, {Akamatsu}, {Gu}, {Kosec}, {Mao}, {Pinto}, {Reiprich}, {Sanders}, {Simionescu}, \& {Werner}}]{Mernier2017}
{Mernier}, F., {de Plaa}, J., {Kaastra}, J.~S., {et~al.} 2017, \aap, 603, A80

\bibitem[{{Mulroy} {et~al.}(2019){Mulroy}, {Farahi}, {Evrard}, {Smith}, {Finoguenov}, {O'Donnell}, {Marrone}, {Abdulla}, {Bourdin}, {Carlstrom}, {D{\'e}mocl{\`e}s}, {Haines}, {Martino}, {Mazzotta}, {McGee}, \& {Okabe}}]{Mulroy19}
{Mulroy}, S.~L., {Farahi}, A., {Evrard}, A.~E., {et~al.} 2019, \mnras, 484, 60

\bibitem[{{Oppenheimer} {et~al.}(2021){Oppenheimer}, {Babul}, {Bah{\'e}}, {Butsky}, \& {McCarthy}}]{Oppenheimer2020}
{Oppenheimer}, B.~D., {Babul}, A., {Bah{\'e}}, Y., {Butsky}, I.~S., \& {McCarthy}, I.~G. 2021, Universe, 7, 209

\bibitem[{{Pearson} {et~al.}(2017){Pearson}, {Ponman}, {Norberg}, {Robotham}, {Babul}, {Bower}, {McCarthy}, {Brough}, {Driver}, \& {Pimbblet}}]{Pearson17}
{Pearson}, R.~J., {Ponman}, T.~J., {Norberg}, P., {et~al.} 2017, \mnras, 469, 3489

\bibitem[{{Pillepich} {et~al.}(2019){Pillepich}, {Nelson}, {Springel}, {Pakmor}, {Torrey}, {Weinberger}, {Vogelsberger}, {Marinacci}, {Genel}, {van der Wel}, \& {Hernquist}}]{pillepich19}
{Pillepich}, A., {Nelson}, D., {Springel}, V., {et~al.} 2019, \mnras, 490, 3196

\bibitem[{{Planck Collaboration} {et~al.}(2016){Planck Collaboration}, {Ade}, {Aghanim}, {Arnaud}, {Ashdown}, {Aumont}, {Baccigalupi}, {Banday}, {Barreiro}, {Bartlett}, {Bartolo}, {Battaner}, {Battye}, {Benabed}, {Beno{\^\i}t}, {Benoit-L{\'e}vy}, {Bernard}, {Bersanelli}, {Bielewicz}, {Bock}, {Bonaldi}, {Bonavera}, {Bond}, {Borrill}, {Bouchet}, {Boulanger}, {Bucher}, {Burigana}, {Butler}, {Calabrese}, {Cardoso}, {Catalano}, {Challinor}, {Chamballu}, {Chary}, {Chiang}, {Chluba}, {Christensen}, {Church}, {Clements}, {Colombi}, {Colombo}, {Combet}, {Coulais}, {Crill}, {Curto}, {Cuttaia}, {Danese}, {Davies}, {Davis}, {de Bernardis}, {de Rosa}, {de Zotti}, {Delabrouille}, {D{\'e}sert}, {Di Valentino}, {Dickinson}, {Diego}, {Dolag}, {Dole}, {Donzelli}, {Dor{\'e}}, {Douspis}, {Ducout}, {Dunkley}, {Dupac}, {Efstathiou}, {Elsner}, {En{\ss}lin}, {Eriksen}, {Farhang}, {Fergusson}, {Finelli}, {Forni}, {Frailis}, {Fraisse}, {Franceschi}, {Frejsel}, {Galeotta}, {Galli}, {Ganga}, {Gauthier}, {Gerbino}, {Ghosh}, {Giard},
  {Giraud-H{\'e}raud}, {Giusarma}, {Gjerl{\o}w}, {Gonz{\'a}lez-Nuevo}, {G{\'o}rski}, {Gratton}, {Gregorio}, {Gruppuso}, {Gudmundsson}, {Hamann}, {Hansen}, {Hanson}, {Harrison}, {Helou}, {Henrot-Versill{\'e}}, {Hern{\'a}ndez-Monteagudo}, {Herranz}, {Hildebrandt}, {Hivon}, {Hobson}, {Holmes}, {Hornstrup}, {Hovest}, {Huang}, {Huffenberger}, {Hurier}, {Jaffe}, {Jaffe}, {Jones}, {Juvela}, {Keih{\"a}nen}, {Keskitalo}, {Kisner}, {Kneissl}, {Knoche}, {Knox}, {Kunz}, {Kurki-Suonio}, {Lagache}, {L{\"a}hteenm{\"a}ki}, {Lamarre}, {Lasenby}, {Lattanzi}, {Lawrence}, {Leahy}, {Leonardi}, {Lesgourgues}, {Levrier}, {Lewis}, {Liguori}, {Lilje}, {Linden-V{\o}rnle}, {L{\'o}pez-Caniego}, {Lubin}, {Mac{\'\i}as-P{\'e}rez}, {Maggio}, {Maino}, {Mandolesi}, {Mangilli}, {Marchini}, {Maris}, {Martin}, {Martinelli}, {Mart{\'\i}nez-Gonz{\'a}lez}, {Masi}, {Matarrese}, {McGehee}, {Meinhold}, {Melchiorri}, {Melin}, {Mendes}, {Mennella}, {Migliaccio}, {Millea}, {Mitra}, {Miville-Desch{\^e}nes}, {Moneti}, {Montier}, {Morgante}, {Mortlock},
  {Moss}, {Munshi}, {Murphy}, {Naselsky}, {Nati}, {Natoli}, {Netterfield}, {N{\o}rgaard-Nielsen}, {Noviello}, {Novikov}, {Novikov}, {Oxborrow}, {Paci}, {Pagano}, {Pajot}, {Paladini}, {Paoletti}, {Partridge}, {Pasian}, {Patanchon}, {Pearson}, {Perdereau}, {Perotto}, {Perrotta}, {Pettorino}, {Piacentini}, {Piat}, {Pierpaoli}, {Pietrobon}, {Plaszczynski}, {Pointecouteau}, {Polenta}, {Popa}, {Pratt}, {Pr{\'e}zeau}, {Prunet}, {Puget}, {Rachen}, {Reach}, {Rebolo}, {Reinecke}, {Remazeilles}, {Renault}, {Renzi}, {Ristorcelli}, {Rocha}, {Rosset}, {Rossetti}, {Roudier}, {Rouill{\'e} d'Orfeuil}, {Rowan-Robinson}, {Rubi{\~n}o-Mart{\'\i}n}, {Rusholme}, {Said}, {Salvatelli}, {Salvati}, {Sandri}, {Santos}, {Savelainen}, {Savini}, {Scott}, {Seiffert}, {Serra}, {Shellard}, {Spencer}, {Spinelli}, {Stolyarov}, {Stompor}, {Sudiwala}, {Sunyaev}, {Sutton}, {Suur-Uski}, {Sygnet}, {Tauber}, {Terenzi}, {Toffolatti}, {Tomasi}, {Tristram}, {Trombetti}, {Tucci}, {Tuovinen}, {T{\"u}rler}, {Umana}, {Valenziano}, {Valiviita}, {Van Tent},
  {Vielva}, {Villa}, {Wade}, {Wandelt}, {Wehus}, {White}, {White}, {Wilkinson}, {Yvon}, {Zacchei}, \& {Zonca}}]{Planck2016}
{Planck Collaboration}, {Ade}, P.~A.~R., {Aghanim}, N., {et~al.} 2016, \aap, 594, A13

\bibitem[{{Popesso} {et~al.}(2005){Popesso}, {Biviano}, {B{\"o}hringer}, {Romaniello}, \& {Voges}}]{Popesso2005}
{Popesso}, P., {Biviano}, A., {B{\"o}hringer}, H., {Romaniello}, M., \& {Voges}, W. 2005, \aap, 433, 431

\bibitem[{{Popesso} {et~al.}(2024{\natexlab{a}}){Popesso}, {Biviano}, {Bulbul}, {Merloni}, {Comparat}, {Clerc}, {Igo}, {Liu}, {Driver}, {Salvato}, {Brusa}, {Bahar}, {Malavasi}, {Ghirardini}, {Robotham}, {Liske}, \& {Grandis}}]{Popesso24}
{Popesso}, P., {Biviano}, A., {Bulbul}, E., {et~al.} 2024{\natexlab{a}}, \mnras, 527, 895

\bibitem[{{Popesso} {et~al.}(2019){Popesso}, {Concas}, {Morselli}, {Schreiber}, {Rodighiero}, {Cresci}, {Belli}, {Erfanianfar}, {Mancini}, {Inami}, {Dickinson}, {Ilbert}, {Pannella}, \& {Elbaz}}]{popesso2019}
{Popesso}, P., {Concas}, A., {Morselli}, L., {et~al.} 2019, \mnras, 483, 3213

\bibitem[{{Popesso} {et~al.}(2024{\natexlab{b}}){Popesso}, {Marini}, \& {Dolag}}]{Popesso2024c}
{Popesso}, P., {Marini}, I., \& {Dolag}, K. 2024{\natexlab{b}}, A\&A submitted, arXiv:2411.17120

\bibitem[{{Powell} {et~al.}(2022){Powell}, {Allen}, {Caglar}, {Cappelluti}, {Harrison}, {Irving}, {Koss}, {Mantz}, {Oh}, {Ricci}, {Shaper}, {Stern}, {Trakhtenbrot}, {Urry}, \& {Wong}}]{powel2022}
{Powell}, M.~C., {Allen}, S.~W., {Caglar}, T., {et~al.} 2022, \apj, 938, 77

\bibitem[{{Pratt} {et~al.}(2010){Pratt}, {Arnaud}, {Piffaretti}, {B{\"o}hringer}, {Ponman}, {Croston}, {Voit}, {Borgani}, \& {Bower}}]{Pratt2010}
{Pratt}, G.~W., {Arnaud}, M., {Piffaretti}, R., {et~al.} 2010, \aap, 511, A85

\bibitem[{{Pratt} {et~al.}(2009){Pratt}, {Croston}, {Arnaud}, \& {B{\"o}hringer}}]{pratt09}
{Pratt}, G.~W., {Croston}, J.~H., {Arnaud}, M., \& {B{\"o}hringer}, H. 2009, \aap, 498, 361

\bibitem[{{Predehl} {et~al.}(2021){Predehl}, {Andritschke}, {Arefiev}, {Babyshkin}, {Batanov}, {Becker}, {B{\"o}hringer}, {Bogomolov}, {Boller}, {Borm}, {Bornemann}, {Br{\"a}uninger}, {Br{\"u}ggen}, {Brunner}, {Brusa}, {Bulbul}, {Buntov}, {Burwitz}, {Burkert}, {Clerc}, {Churazov}, {Coutinho}, {Dauser}, {Dennerl}, {Doroshenko}, {Eder}, {Emberger}, {Eraerds}, {Finoguenov}, {Freyberg}, {Friedrich}, {Friedrich}, {F{\"u}rmetz}, {Georgakakis}, {Gilfanov}, {Granato}, {Grossberger}, {Gueguen}, {Gureev}, {Haberl}, {H{\"a}lker}, {Hartner}, {Hasinger}, {Huber}, {Ji}, {Kienlin}, {Kink}, {Korotkov}, {Kreykenbohm}, {Lamer}, {Lomakin}, {Lapshov}, {Liu}, {Maitra}, {Meidinger}, {Menz}, {Merloni}, {Mernik}, {Mican}, {Mohr}, {M{\"u}ller}, {Nandra}, {Nazarov}, {Pacaud}, {Pavlinsky}, {Perinati}, {Pfeffermann}, {Pietschner}, {Ramos-Ceja}, {Rau}, {Reiffers}, {Reiprich}, {Robrade}, {Salvato}, {Sanders}, {Santangelo}, {Sasaki}, {Scheuerle}, {Schmid}, {Schmitt}, {Schwope}, {Shirshakov}, {Steinmetz}, {Stewart}, {Str{\"u}der},
  {Sunyaev}, {Tenzer}, {Tiedemann}, {Tr{\"u}mper}, {Voron}, {Weber}, {Wilms}, \& {Yaroshenko}}]{Predehl2021}
{Predehl}, P., {Andritschke}, R., {Arefiev}, V., {et~al.} 2021, \aap, 647, A1

\bibitem[{{Ragagnin} {et~al.}(2022){Ragagnin}, {Andreon}, \& {Puddu}}]{Ragagnin22}
{Ragagnin}, A., {Andreon}, S., \& {Puddu}, E. 2022, \aap, 666, A22

\bibitem[{{Rasmussen} \& {Ponman}(2009)}]{Rasmussen09}
{Rasmussen}, J. \& {Ponman}, T.~J. 2009, \mnras, 399, 239

\bibitem[{{Robotham} {et~al.}(2011){Robotham}, {Norberg}, {Driver}, {Baldry}, {Bamford}, {Hopkins}, {Liske}, {Loveday}, {Merson}, {Peacock}, {Brough}, {Cameron}, {Conselice}, {Croom}, {Frenk}, {Gunawardhana}, {Hill}, {Jones}, {Kelvin}, {Kuijken}, {Nichol}, {Parkinson}, {Pimbblet}, {Phillipps}, {Popescu}, {Prescott}, {Sharp}, {Sutherland}, {Taylor}, {Thomas}, {Tuffs}, {van Kampen}, \& {Wijesinghe}}]{Robotham2011}
{Robotham}, A.~S.~G., {Norberg}, P., {Driver}, S.~P., {et~al.} 2011, \mnras, 416, 2640

\bibitem[{{Rozo} {et~al.}(2009){Rozo}, {Rykoff}, {Evrard}, {Becker}, {McKay}, {Wechsler}, {Koester}, {Hao}, {Hansen}, {Sheldon}, {Johnston}, {Annis}, \& {Frieman}}]{rozo2009}
{Rozo}, E., {Rykoff}, E.~S., {Evrard}, A., {et~al.} 2009, \apj, 699, 768

\bibitem[{{Schaye} {et~al.}(2015){Schaye}, {Crain}, {Bower}, {Furlong}, {Schaller}, {Theuns}, {Dalla Vecchia}, {Frenk}, {McCarthy}, {Helly}, {Jenkins}, {Rosas-Guevara}, {White}, {Baes}, {Booth}, {Camps}, {Navarro}, {Qu}, {Rahmati}, {Sawala}, {Thomas}, \& {Trayford}}]{Schaye2015}
{Schaye}, J., {Crain}, R.~A., {Bower}, R.~G., {et~al.} 2015, \mnras, 446, 521

\bibitem[{{Schaye} {et~al.}(2023){Schaye}, {Kugel}, {Schaller}, {Helly}, {Braspenning}, {Elbers}, {McCarthy}, {van Daalen}, {Vandenbroucke}, {Frenk}, {Kwan}, {Salcido}, {Bah{\'e}}, {Borrow}, {Chaikin}, {Hahn}, {Hu{\v{s}}ko}, {Jenkins}, {Lacey}, \& {Nobels}}]{Schaye2023}
{Schaye}, J., {Kugel}, R., {Schaller}, M., {et~al.} 2023, \mnras, 526, 4978

\bibitem[{{Seppi} {et~al.}(2022){Seppi}, {Comparat}, {Bulbul}, {Nandra}, {Merloni}, {Clerc}, {Liu}, {Ghirardini}, {Liu}, {Salvato}, {Sanders}, {Wilms}, {Dwelly}, {Dauser}, {K{\"o}nig}, {Ramos-Ceja}, {Garrel}, \& {Reiprich}}]{Seppi2022}
{Seppi}, R., {Comparat}, J., {Bulbul}, E., {et~al.} 2022, \aap, 665, A78

\bibitem[{Springel \& Hernquist(2003)}]{springel_cosmological_2003}
Springel, V. \& Hernquist, L. 2003, MNRAS, 339, 289

\bibitem[{Springel {et~al.}(2001)Springel, White, Tormen, \& Kauffmann}]{springel_populating_2001}
Springel, V., White, S. D.~M., Tormen, G., \& Kauffmann, G. 2001, MNRAS, 328, 726

\bibitem[{{Sun} {et~al.}(2009){Sun}, {Voit}, {Donahue}, {Jones}, {Forman}, \& {Vikhlinin}}]{Sun2009}
{Sun}, M., {Voit}, G.~M., {Donahue}, M., {et~al.} 2009, \apj, 693, 1142

\bibitem[{{Tempel} {et~al.}(2017){Tempel}, {Tuvikene}, {Kipper}, \& {Libeskind}}]{Tempel2017}
{Tempel}, E., {Tuvikene}, T., {Kipper}, R., \& {Libeskind}, N.~I. 2017, \aap, 602, A100

\bibitem[{{Tinker}(2021)}]{tinker2021}
{Tinker}, J.~L. 2021, \apj, 923, 154

\bibitem[{Tornatore {et~al.}(2007)Tornatore, Borgani, Dolag, \& Matteucci}]{tornatore_chemical_2007}
Tornatore, L., Borgani, S., Dolag, K., \& Matteucci, F. 2007, MNRAS, 382, 1050

\bibitem[{{Vikhlinin} {et~al.}(2006){Vikhlinin}, {Kravtsov}, {Forman}, {Jones}, {Markevitch}, {Murray}, \& {Van Speybroeck}}]{Vikhlinin06}
{Vikhlinin}, A., {Kravtsov}, A., {Forman}, W., {et~al.} 2006, \apj, 640, 691

\bibitem[{{Vladutescu-Zopp} {et~al.}(2023){Vladutescu-Zopp}, {Biffi}, \& {Dolag}}]{VZS23}
{Vladutescu-Zopp}, S., {Biffi}, V., \& {Dolag}, K. 2023, \aap, 669, A34

\bibitem[{Wiersma {et~al.}(2009)Wiersma, Schaye, \& Smith}]{wiersma_effect_2009}
Wiersma, R. P.~C., Schaye, J., \& Smith, B.~D. 2009, MNRAS, 393, 99

\bibitem[{Yang {et~al.}(2005)Yang, Mo, van~den Bosch, Weinmann, Li, \& Jing}]{Yang2005}
Yang, X., Mo, H.~J., van~den Bosch, F.~C., {et~al.} 2005, Monthly Notices of the Royal Astronomical Society, 362, 711

\bibitem[{{Zhang} {et~al.}(2024){Zhang}, {Comparat}, {Ponti}, {Meloni}, {Nandra}, {Haberl}, {Locatelli}, {Zhang}, {Sanders}, {Zheng}, {Liu}, {Popesso}, {Liu}, {Truong}, {Pillepich}, {Predehl}, \& {Salvato}}]{Zhang24a}
{Zhang}, Y., {Comparat}, J., {Ponti}, G., {et~al.} 2024, arXiv e-prints, arXiv:2401.17308

\end{thebibliography}

\begin{appendix} 
\section{The key role of the halo mass proxy in the stacking}

\begin{figure*}
\includegraphics[width=\textwidth]{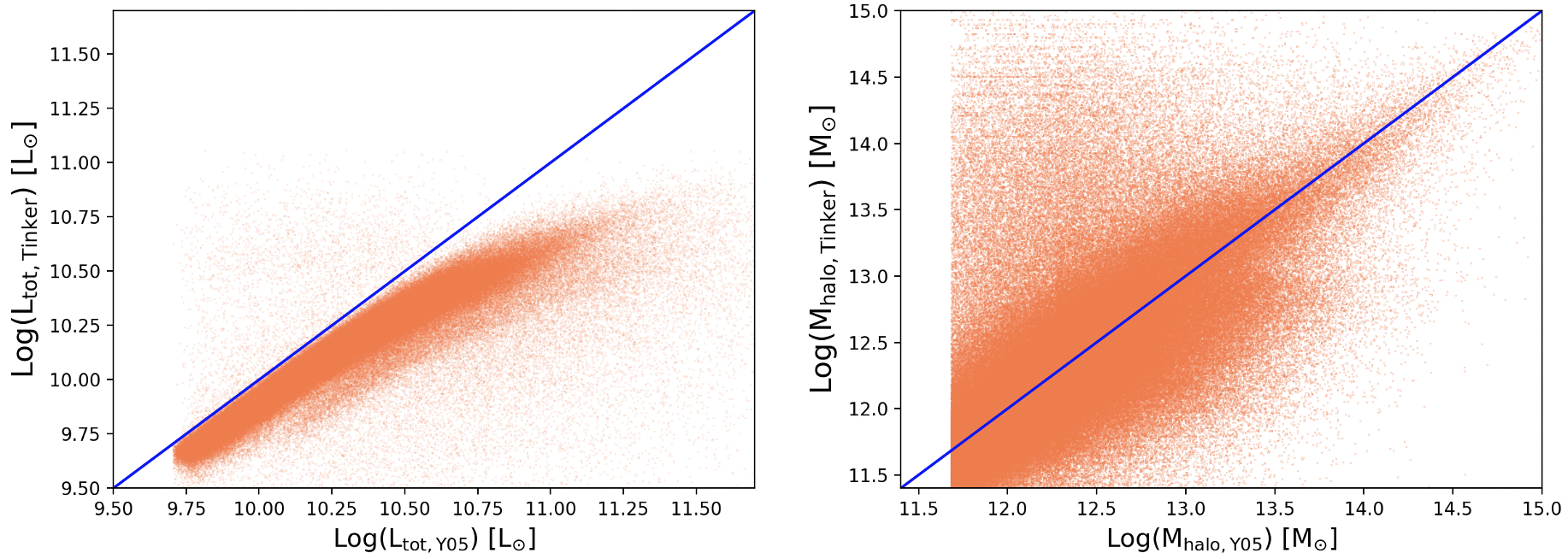}
\caption{{\it{Left panel:}} Comparison between the total optical luminosity provided by T21 and the value provided by Y05. The blue solid line indicates the one-to-one relation. {\it{Right panel}}: Comparison between the halo mass proxy provided by T21 and the value provided by Y05 based on the respective $L_{tot}$. The blue solid line indicates the one-to-one relation.}
\label{tinker}
\end{figure*}

\begin{figure}
\includegraphics[width=0.5\textwidth]{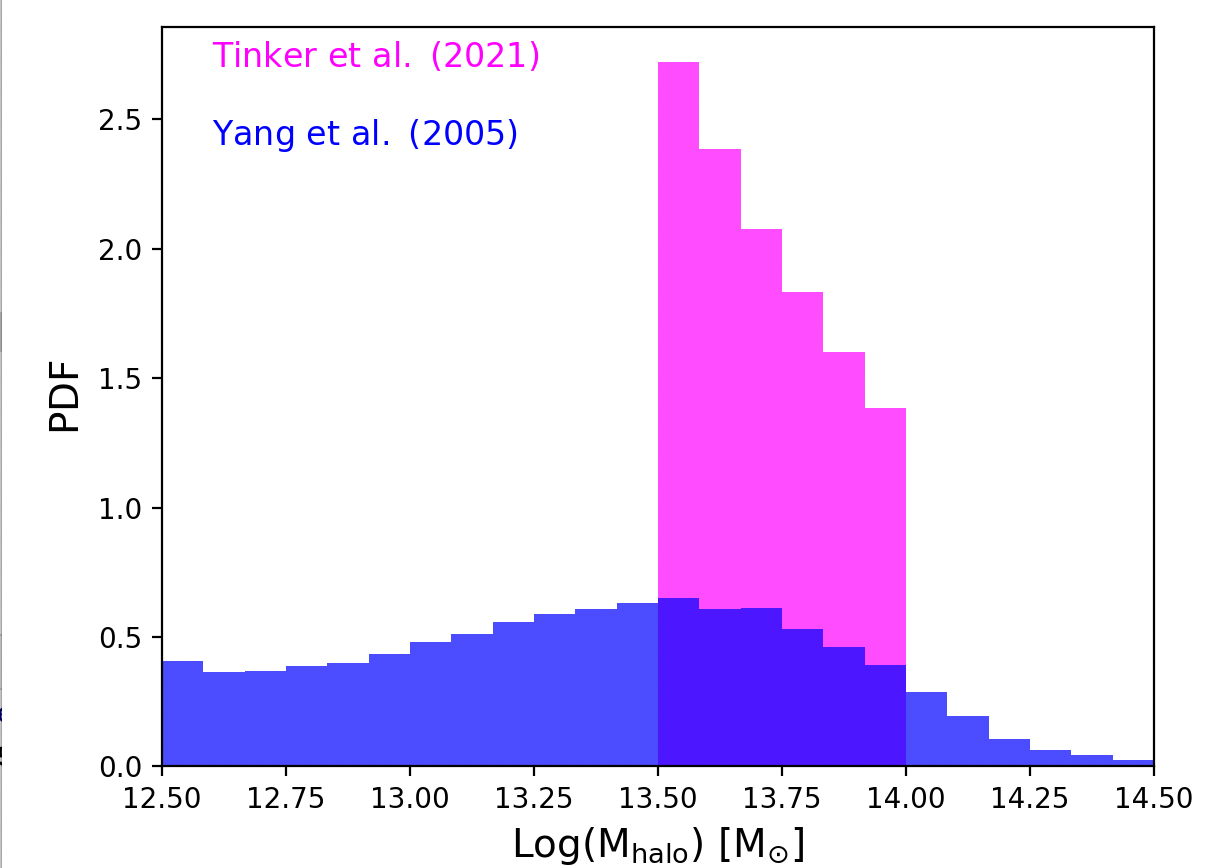}
\caption{Distribution of the halo masses of T21 group catalog in the $10^{13.5}-10^{14}$ $M_{\odot}$ halo mass bin (magenta histogram) and of the matched groups in the Y05 (blue histogram).}
\label{tinker_mass}
\end{figure}

The comparison between Magneticum IGM profile predictions and eROSITA stacked observations of optically selected groups shows substantial differences in the observational data (see Fig. \ref{zhang} and \ref{mio}). This discrepancy appears to stem primarily from the distinct halo mass proxies used in the optical group prior samples.

In particular, we compare the X-ray surface brightness profiles in \citet{Zhang24a} with those of \citet{Popesso2024c}. \citet{Zhang24a} stacks optically selected groups from the T21 sample \citep{tinker2021}, while \citet{Popesso2024c} employs the R11 sample. Additionally, the stacking methods differ; notably, \citet{Zhang24a} applies AGN contamination corrections using the \cite{Comparat2022} model and subtracts satellite X-ray background (XRB) contributions, although satellite galaxies within groups are generally quiescent \citep{popesso2019}. However, these adjustments do not seem to account for the observed discrepancies. Instead, the primary issue appears to be the X-ray emission levels in stacks prior to AGN and XRB corrections, which are consistently low. This likely results from differences in halo mass proxies, as well as potential contamination and completeness issues in the T21 prior catalog, not discussed in \citet{Zhang24a} after further galaxy selection refinements.

To explore this, we matched the T21 group catalog with the Y05 sample in the SDSS coverage area. Since the R11 catalog used by \citet{Popesso2024c} overlaps minimally with T21 in sky area and redshift, we focus on comparing T21 and Y05. We matched group samples by aligning the central galaxies identified by each algorithm. In 87\% of cases, we found correspondence between T21 and Y05 group centrals, suggesting a shared bulk population. However, in 13\% of cases, central galaxies in one catalog were satellites in the other. This allows a direct comparison of group mass and total optical luminosity ($L_{opt}$) associated with matched central galaxies.

As shown in Fig. \ref{tinker} (right panel), while T21 and R11 mass proxies exhibit a one-to-one relationship, there is significant scatter (0.68 dex). Both proxies depend on total group $L_{opt}$, though definitions differ. For Y05, $L_{opt}$ includes all group members brighter than 19.5 mag in the r-band, including the central galaxy, while T21 calculates $L_{opt}$ from satellite luminosities only. In Fig. \ref{tinker} (left panel), we see deviations from a one-to-one $L_{opt}$ relationship, with T21 values consistently lower than Y05’s, even when central galaxies are included. This likely reflects differing $L_{opt}$ definitions and membership identifications.

The greater scatter in mass proxy relations compared to $L_{opt}$ (0.68 vs. 0.17 dex) indicates that the $L_{opt}$-to-halo mass calibration may also contribute. Y05 employs a single power-law for this conversion, while T21 applies distinct calibrations for blue and red galaxy groups \citep{Alpaslan2020}. Although weak lensing studies suggest a possible dependence on galaxy color \citep[e.g.,][]{Mandelbaum2016}, this trend is absent in local and high-redshift samples, where $L_{opt}$ and halo mass derived from X-ray data show a consistent single power-law relation across galaxy properties \citep{ghazaleh1,ghazaleh2}.

In Fig. \ref{tinker_mass}, we observe that T21’s $10^{13.5}-10^{14} M_{halo}$ mass bin (used in Fig. \ref{zhang}, left panel) spans a broader mass range under the Y05 proxy. The Y05 algorithm’s effectiveness in estimating halo mass and identifying central/member galaxies is supported by studies such as \citet{Marini24b}, suggesting that T21 calibration may admit significant low-mass contamination across mass bins, potentially lowering mean X-ray flux in stacks by including low-luminosity, low-mass objects. A complete comparison would require incorporating the T21 algorithm into the mock dataset, as done for R11, Y05, and T17 in \citet{Marini24b}.

In conclusion, our analysis highlights that stacked X-ray results are highly sensitive to halo mass proxy systematics and selection biases in the prior catalog.

\end{appendix}

\end{document}